\documentclass[12pt]{article}
\title{}
\author{}
\usepackage[dvipdfmx]{graphicx}
\usepackage{comment}
\usepackage{epsf}
\usepackage{amsmath}
\usepackage{graphics}
\usepackage{amsfonts}
\usepackage{amssymb}
\usepackage{latexsym}
\usepackage{color}
\usepackage{cite}
\input{colordvi.tex}

\usepackage{feynmp}
\renewcommand{\thefootnote}{\#\arabic{footnote}}

\newcommand{\be}{\begin{equation}}
\newcommand{\ee}{\end{equation}}
\newcommand{\bea}{\begin{eqnarray}}
\newcommand{\eea}{\end{eqnarray}}

\def\m{\mathrm}
\def\be{\begin{equation}}
\def\ee{\end{equation}}

\def\al{\alpha}
\def\be{\begin{equation}}
\def\ee{\end{equation}}

\def\fr{\frac}

\def\MS{M_\odot}
\def\be{\begin{equation}}
\def\ee{\end{equation}}

\def\fr{\frac}

\def\Om{\Omega}
\def\al{\alpha}
\def\si{\sigma}

\def\m{\mathrm}

\def\ra{r_\m{anni}}
\def\rK{r_\m{Kep}}
\def\MP{M_\m{PBH}}
\def\MU{M_\m{UCMH}}

\def\zt{z_\m{turn}}

\usepackage{color}
\date{\today}
\begin{document}
\setcounter{footnote}{0}

\begin{titlepage}
\begin{flushright}
RESCEU-9/14, KEK-TH-1736, KEK-Cosmo-146
\end{flushright}
\begin{center}

\vskip .5in

{\Large \bf
Testing scenarios of primordial black holes \\being the seeds of supermassive black holes \\by
ultracompact minihalos and CMB $\mu$-distortions
}

\vskip .45in

{\large
Kazunori Kohri$^{1}$,
Tomohiro Nakama$^{2,3}$
and 
Teruaki Suyama$^{3}$
}

\vskip .45in%

{\em 
$^1$
  Cosmophysics Group, Theory Centre, IPNS, KEK, Tsukuba 305-0801, Japan, \\
  and The Graduate University for Advanced Study (Sokendai), Tsukuba
305-0801, Japan
}\\
{\em
$^2$
  Department of Physics, Graduate School of Science,\\ The University of Tokyo, Tokyo 113-0033, Japan
}\\
{\em
$^3$
  Research Center for the Early Universe (RESCEU), Graduate School
  of Science,\\ The University of Tokyo, Tokyo 113-0033, Japan
}

\end{center}

\vskip .4in

\begin{abstract}
Supermassive black holes and intermediate mass black holes are believed to exist in the Universe.
There is no established astrophysical explanation for their origin and 
considerations have been made in the literature that 
those massive black holes (MBHs) may be primordial black holes (PBHs), 
black holes which are formed in the early universe (well before the matter-radiation equality) 
due to the direct collapse of primordial overdensities.
This paper aims at discussing the possibility of excluding the PBH scenario
as the origin of the MBHs.
We first revisit the constraints on PBHs obtained from the CMB distortion that the seed density perturbation causes.
By adopting a recent computation of the CMB distortion sourced by the seed density perturbation
and the stronger constraint on the CMB distortion set by the COBE/FIRAS experiment
used in the literature,
we find that PBHs in the mass range 
$6\times 10^4~M_\odot \sim 5 \times 10^{13}~M_\odot$
are excluded.
Since PBHs lighter than $6 \times 10^4~M_\odot$ are not excluded from the
non-observation of the CMB distortion,
we propose a new method which can potentially exclude smaller PBHs as well.
Based on the observation that large density perturbations required to create PBHs
also result in the copious production of ultracompact minihalos (UCMHs),
compact dark matter halos formed at around the recombination,
we show that weakly interacting massive particles (WIMPs) as dark matter
annihilate efficiently inside UCMHs
to yield cosmic rays far exceeding the observed flux.
Our bound gives severe restriction on the compatibility between the particle
physics models for WIMPs and the PBH scenario as the explanation of MBHs.
\end{abstract}
\end{titlepage}

\renewcommand{\thepage}{\arabic{page}}
\setcounter{page}{1}
\renewcommand{\thefootnote}{\#\arabic{footnote}}

\section{Introduction}
Supermassive black holes (SMBHs) ranging from $10^6~{\rm M_\odot}$ to $3\times 10^9~{\rm M_\odot}$
are believed to reside at the center of many galaxies including the Milky Way \cite{Kormendy:1995er}. 
They are thought to be descendants of seed BHs (observed as quasars) at high redshifts ($z \sim 3$). 
Those quasars are likely to have undergone mergers associated with galaxy mergers and 
have grown to become the observed SMBHs.
Furthermore, SMBHs are also observed at higher redshifts ($z =6\sim 7$) \cite{Fan:2003wd,Mortlock:2011va}.
Quite intriguingly, some of them have masses exceeding $10^9~{\rm M_\odot}$,
which makes astrophysical explanations of their origin challenging. 
In addition, smaller mass BHs, the so-called Intermediate-mass BHs (IMBHs), 
ranging between $10^2~{\rm M_\odot}$ to $10^6~{\rm M_\odot}$ probably explain observed
luminous X-ray sources \cite{Maccarone:2007dd,Farrell:2010bf}.
Despite the observational revelation of the ubiquitous existence
of these massive BHs (MBHs) \footnote{In this paper, we use the massive black holes (MBHs)
to refer to BHs whose mass is larger than the solar mass.}, 
the established theoretical explanation of their origin is still missing.

The observed MBHs, either SMBHs or IMBHs, or even both, are possibly primordial black holes (PBHs) that
formed by the direct gravitational collapse of density perturbations having 
extremely large amplitude ($\delta ={\cal O}(1)$) when the Universe
is still dominated by radiation\cite{Yokoyama:1995ex,Duechting:2004dk,Kawasaki:2012kn,Kawasaki:2012wr,Kohri:2012yw}.
PBHs are formed right after the coherent length of large inhomogeneity which was originally 
super-Hubble length scale becomes comparable to 
the Hubble horizon length \cite{Hawking:1971ei,Carr:1974nx,Carr:1975qj}.
The mass of the resulting PBH is approximately equal to the horizon mass, 
the radiation energy density multiplied by the Hubble volume at the time of the PBH formation.
Depending on when PBHs are formed (in other words, depending on the comoving coherent length 
of density perturbation), the mass of PBHs covers a vast range, 
starting from the Planck mass $\sim 10^{-5}~{\rm g}$ up to $5 \times 10^{17}~{\rm M_\odot}$
(for detailed investigation of cosmological consequences of PBHs for a wide range of masses,
see \cite{Carr:2009jm}).
The shorter the comoving coherent length is, the smaller the mass of the PBH is.
The characteristic feature of the PBH scenario is that producing BHs as massive as
$10^9~{\rm M_\odot}$ is easily achieved by simply preparing primordial perturbation 
with sufficiently large amplitude with suitable comoving coherent length.
Once such density perturbation is prepared, the PBH formation inevitably occurs. 
The drawback of the PBH scenario is that realizing such large amplitude of density
perturbations heavily relies on inflation models which we do not know.
Several papers have appeared in which some inflationary models are contrived to
produce PBHs for the purpose of explaining the observed 
MBHs \cite{Yokoyama:1995ex,Kohri:2007qn,Kawasaki:2012kn,Kawasaki:2012wr,Alabidi:2012ex}. 
However, it is difficult to conclude or even to test if the PBH scenario is indeed correct or not.

The aim of this paper is to investigate if it is possible to exclude the PBH scenario
as the origin of the MBHs.
In this paper, we focus on two astrophysical effects that PBH formation leads to.
The first effect is the distortion of black body spectrum of the Cosmic microwave background (CMB)
caused by the Silk damping of the density perturbations.
Although the amplitude of the density perturbation for the PBH formation typically corresponds to the very tail
of the distribution function, the standard deviation is still large enough and leads to the
observable CMB distortion for some range of PBH mass.
Given the non-detection of the CMB black body distortion, 
PBH in such mass range is excluded.
Note that this idea itself is not new. 
Actually in \cite{Carr:1993aq,Carr:1994ar}
it is applied, with the combined use of the COBE data, to exclude PBHs in some mass range.
By adopting a modern computation of the CMB distortion and
stronger observational bounds \cite{Fixsen:1996nj} than were used in \cite{Carr:1993aq,Carr:1994ar}, 
PBHs with their mass larger than $10^5~M_\odot$ were excluded in \cite{Chluba:2012we}
for the locally scale invariant power spectrum.
In the next section, we will briefly revisit this issue by adopting a simple $\delta$-function 
shape for the primordial power spectrum, which is used throughout this paper,
and find exclusion range of the PBH mass similar to that of \cite{Chluba:2012we}.
The exclusion of PBH mass $\gtrsim 10^5~M_\odot$ as seeds of the SMBHs may be thought to disfavor, 
but not preclude, the PBH scenario as the origin of the SMBHs existing at 
high redshift $\gtrsim 6$ because of the short available time to increase the PBH
mass from less than $10^5~M_\odot$ to more than $10^9~M_\odot$ by $z =6\sim 7$.
We need efficient accretion in order to achieve such BH growth and it is not clear
if it is really possible from the astrophysical point of view.
On the other hand, PBHs with mass less than $10^5~M_\odot$ are in favor of the present SMBHs
and the IMBHs.
For instance, it was demonstrated in \cite{Bean:2002kx} that PBHs lighter than 
$10^5~M_\odot$ can evolve into the present SMBHs by mergers.
Thus, in addition to the exclusion of PBHs heavier than $10^5~M_\odot$,
it is equally important to exclude PBHs lighter than $10^5~M_\odot$.
This can be potentially possible by making use of the second effect we consider in the following.

The effect is the emission of high energy cosmic rays due to annihilations of 
dark matter particles in compact dark matter halos.
Once more, PBHs form at sites where the density perturbation is as large as ${\cal O}(1)$ \cite{Carr:1974nx}.
The ${\cal O}(1)$ amplitude of the density perturbation corresponds to the tail of
the Gaussian probability distribution function, typically taken to be $10~\sigma$ deviation
to be consistent with observational upper bounds on the PBH abundance \cite{Carr:2009jm},
where $\sigma^2$ is the variance of $\delta$. 
This implies that sites of PBH formation are extremely sparse and 
the amplitude of the density perturbation at most places in the Universe takes 
$\delta \sim \sigma \simeq {\cal O}(10^{-2})$, which is not enough for PBH formation but is still
considerably larger than the density perturbation observed at CMB scales. 
As was pointed out by Ricotti and Gould \cite{Ricotti:2009bs}, 
density perturbations on small scales having amplitude larger than $\sim 10^{-3}$, 
a value which is much smaller than the typical amplitude of ${\cal O}(10^{-2})$ mentioned above, 
lead to gravitational collapse of dark matter (DM) at around the time of the recombination
to form what are called ultracompact minihalos (UCMHs) in the literature. 
It would be worthwhile to mention that the formation of these UCMHs is due to the combined effects of 
the slow growth during the radiation-domination
\footnote{We consider density perturbations which re-enter the Hubble horizon
after the kinetic decoupling of the dark matter particles.
As we show later, this places a lower bound on the mass of the PBHs for which our 
constraints by the UCMHs can be applied.
} and the relatively rapid growth during the 
matter domination, the growth in proportion to the scale factor. 
This suggests that the formation of a numerous number of UCMHs is an inevitable outcome of the PBH scenario.
If DM particles are discovered by colliders such as Large Hadron Collider (LHC) 
or any other direct experiments in the future and in addition the interaction strengths
with standard model (SM) particles are measured,
which is a plausible possibility from the particle physics point of view,
UCMHs are efficient factories emitting cosmic rays out of DM annihilations
mostly occurring at the core of the UCMHs where the DM density is highest 
\cite{Scott:2009tu,Lacki:2010zf}.
Among the final annihilation yields are, for instance, high energy photons
which can be observed by the cosmic-ray detectors such as Fermi-LAT.
The possibility then arises that non-observation of such DM signals coming from unidentified 
sources is inconsistent with the expected intensity of cosmic rays coming from the UCMHs, 
estimated based on the DM properties potentially measured by terrestrial experiments 
and on the assumption that PBHs are the origin of the MBHs. 
In this case we can falsify this assumption.
It should be noted that a similar logic can be used to constrain the power
of small scale perturbations and in fact intensive investigations have been done in the literature
\cite{Berezinsky:2010kq,Josan:2010vn,Yang:2011jb,Bringmann:2011ut,Yang:2011eg,Li:2012qha,Yang:2012qi,Yang:2013dsa}. 

Based on this consideration, we aim to make a proposal which can 
test the PBH scenario as the origin of the MBHs by assessing the maximum strengths 
of interactions between DM particles and the SM ones allowed by the recent observations
of cosmic-rays assuming PBHs explain MBHs.
It will turn out that our proposal is effective for WIMP dark matter.
As for the cosmic-ray experiments, we use the results of Fermi-LAT which measures
the gamma-rays with good angular and energy resolution. 
We also use the atmospheric neutrino flux to constrain DM properties. 
We consider three processes $\chi \chi \to b{\bar b}$, $\chi \chi \to W^+W^-$ and $\chi \chi \to \tau^+\tau^-$,
where $\chi$ denotes a DM particle, as typical annihilation modes and we place limits 
on each cross section, assuming only one mode dominates the others for each case.
The conversion from the intermediate yields such as bottom quarks or W bosons 
to $\gamma -$rays is calculated by using the public code PYTHIA\cite{Sjostrand:2006za}.
We find that if the observed MBHs are of PBH origin, 
the upper bound on each cross section is mostly a lot less than so called canonical value, 
$\langle \sigma v \rangle_{\rm can} = 3 \times 10^{-26}~{\rm cm}^3{\rm s}^{-1}$, with the precise
bound depending on the PBHs mass as well as density profile of the UCMHs.
Given that WIMPs such as neutralino appearing in many extended SM models typically have 
the cross section of order the canonical value unless some fine-tuning is imposed,
our bounds show that the PBH scenario is strongly disfavored if the WIMPs having
the typical interaction strengths with the SM particles are discovered in future.

\section{Constraints on the abundance of PBHs obtained from CMB $\mu$-distortion}
\label{sec-mu}

\begin{figure}[t]
\begin{center}
\includegraphics[width=13cm,keepaspectratio,clip]{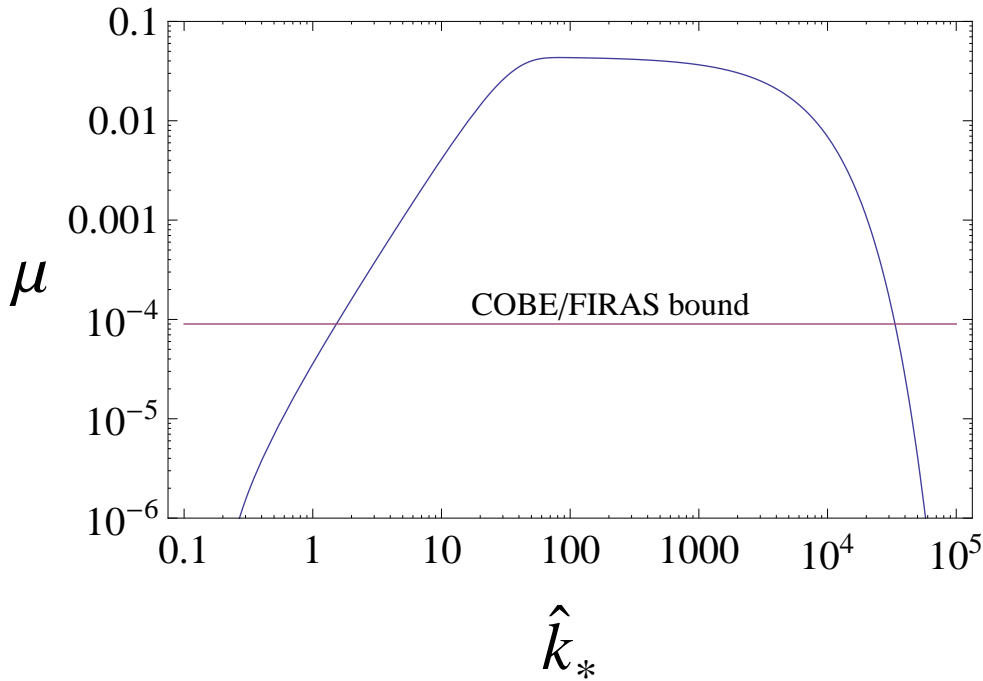}
\end{center}
\caption{
The $\mu$ distortion generated from a single $k$-mode, with $A_{\zeta}=0.02$. 
The horizontal line corresponds to the $2\sigma$ upper limit provided by 
COBE/FIRAS. 
}
\label{mu}
\end{figure}

Constraints on the abundance of PBHs obtained from CMB distortions 
were investigated in 1993 and 1994 by Carr et al.\cite{Carr:1993aq,Carr:1994ar}, 
and in Chluba et al. in 2012 \cite{Chluba:2012we} (here after CEB) the upper bound on the amplitude of the primordial power spectrum 
was derived for the locally scale invariant spectrum with a Gaussian filter, using CMB distortions.
The obtained upper bound is a few orders of magnitude smaller than the
PBH bound around $M_{\rm PBH} \simeq 6 \times 10^4~M_\odot$, 
implying a severe restriction on the abundance of such PBHs.
In this section, following CEB, we briefly revisit this issue by considering $\delta$-function
shape of the primordial power spectrum which is used
throughout this paper
\footnote
{
In typical models (e.g., \cite{Kawasaki:2012kn}) predicting the formation of PBHs as the seeds of SMBHs, 
the power spectrum of curvature perturbation has a sharp peak, 
the height of which exceeds ${\cal O}(0.01)$. 
The width of the peak should be finite but it cannot be arbitrarily wide to avoid overproduction of PBHs 
of masses irrelevant to the seeds of SMBHs. 
In addition, if we take into account the effects of the finite width of the spectrum, it leads to more production of $\mu$ distortion 
since in this case more than one $k$-modes contribute to $\mu$ distortion. For example, if we consider a 
step-like power spectrum $\Delta P_{\zeta}(k)=2\pi^2A_\zeta k^{-3}(1\mathrm{Mpc}^{-1}<k), 0(\rm{otherwise})$, 
the resultant $\mu$ distortion is $\mu\sim 11A_\zeta$ (see CEB). 
Similarly, more UCMHs should be formed if we assume a wider power spectrum 
and so estimations based on one single $k-$mode obtained in this paper provide conservative upper bounds on the scenario of PBHs as the seeds of SMBHs. 
Therefore, for our purposes it is sufficient to restrict our attention to a delta-function like power spectrum. 
}. 
In CEB, the primordial power spectrum of curvature perturbation was decomposed as follows;
\begin{equation}
P_\zeta=P^{\rm{st}}_\zeta(k)+\Delta P_\zeta(k),
\end{equation}
where the first term represents the standard almost scale-invariant power spectrum,
which has been determined by CMB experiments accurately, 
with the second term denoting the deviation from this standard spectrum.
Let us consider the $\delta$-function like $\Delta P_\zeta(k)$ parameterized as follows:
\begin{equation}
\Delta P_\zeta(k)=2\pi^2A_\zeta k^{-2}\delta(k-k_*).
\end{equation}
It turns out that the $\mu$ distortion originating from a single $k$-mode 
is approximated by
\begin{equation}
\mu\sim 2.2A_\zeta
\left[
\exp\left(-\frac{\hat{k}_*}{5400}\right)
-\exp\left(-\left[\frac{\hat{k}_*}{31.6}\right]^2\right)
\right],
\end{equation}
where $k_*=\hat{k}_*\mathrm{Mpc}^{-1}$. 
The COBE/FIRAS experiment provides the 2$\sigma$ upper limit as 
$\mu\lesssim 9\times 10^{-5}$ \cite{Fixsen:1996nj}.
Noting that $A_\zeta \gtrsim {\cal O}(0.01)$ is necessary to produce 
PBHs to an observationally relevant level \cite{Josan:2009qn},
we can plot $\mu$ as a function of ${\hat k}_*$.
Fig. \ref{mu} shows the plot of $\mu$ with $A_\zeta$ fixed to $0.02$.
We find that any perturbation mode in a range 
$1 \lesssim {\hat k}_* \lesssim 3\times 10^4$ produces $\mu$ larger than the
COBE/FIRAS upper bound.
Therefore, PBHs formed from the density perturbation in the above ${\hat k}_*$ range are excluded. 
This conclusion is insensitive to the change of $A_\zeta$ (as long as it is ${\cal O}(0.01)$), 
as is evident from the figure.
Since $k_*$ is related to the PBH mass (see Eq.~(\ref{pbh-peak-mass})),
the above ${\hat k}_*$ range can be translated into the PBH mass range as
$6\times 10^4~M_\odot \lesssim M_{\rm PBH} \lesssim 5 \times 10^{13}~M_\odot$
\footnote{Density perturbations corresponding to larger PBH masses 
generate $y$-type distortion which is also constrained by the COBE/FIRAS experiment.
From the view point of the observed supermassive black holes,
such PBHs are too heavy and we do not consider this case in this paper.
}.
PBHs in this mass range are basically ruled out. 
In the next section, we predominantly focus on PBHs whose mass is at most $5\times 10^4M_{\odot}$ as the 
seeds of the MBHs, which evade the constraints from the $\mu$ distortion. 
\section{Cosmic rays from the UCMHs}
As explained in the Introduction, the production of PBHs requires large amplitude of 
density perturbation.
Such density perturbation also entails the formation of UCMHs.
Since the density inside the UCMHs is large, we expect a considerable amount of 
cosmic rays due to annihilation of dark matter particles if they annihilate.
From the non-observation of such cosmic rays, we can place upper bounds on 
dark matter annihilation cross sections assuming that at least part of the observed 
MBHs are the PBHs, which is the purpose of this section.

The mass distribution of the observed MBHs \cite{Kelly:2011ab} implies that the initial mass function of the seed 
PBHs has non-trivial shape, and hence the initial power spectrum of the primordial perturbation does too.
However, complex natures of the PBH evolution due to mergers and accretions defeat
precise mapping between the mass function of the MBHs and the initial power spectrum.
In this paper, instead of investigating the evolution of PBHs, we simply consider the
monochromatic power spectrum of the density perturbation (see also the footnote 3),
which, as we will see, results in the PBHs having approximately the same mass equal
to the horizon mass evaluated when the density perturbation reenters the Hubble horizon.
Although this may oversimplify the realistic situation, we will show that the upper limits of the dark matter 
annihilation cross sections we obtain 
depend very weakly on $M_{\rm PBH}$, assuming that
all the PBHs have the same mass $M_{\rm PBH}$.
Therefore, even if we take into account the distribution of $M_{\rm PBH}$,
we expect that our constraints would not differ significantly.
In the following, we begin by providing the mass function of PBHs and that of UCMHs to estimate the typical masses under the assumption of 
the monochromatic power spectrum of the density perturbations, 
though in the subsequent sections we assume the monochromatic mass functions for both PBHs and UCMHs.

\subsection{Mass functions of PBHs and UCMHs}
In order to evaluate the mass functions of PBHs and UCMHs, 
we first need to specify the initial power spectrum of the density perturbation.
We consider the monochromatic power spectrum, {\it i.~e.~}
\begin{equation}
P_{\cal R}=2\pi^2 {\cal P_*}k^{-2} \delta (k-k_*), \label{ini-P}
\end{equation}
where ${\cal R}$ is the curvature perturbation on the comoving slice.
In the absence of the isocurvature perturbation, 
${\cal R}$ is conserved on super-Hubble scales.

To connect ${\cal P}_{\cal R}$ with the mass function, 
we follow the formulation given in \cite{Kim:1996hr}.
We first introduce a fixed but artificial (that is, arbitrary) initial time 
$t=t_{\rm ini}$ before the PBH (or UCMH) formation in the comoving slice.
In this slice, the density contrast $\delta$ and ${\cal R}$ is related as
\begin{equation}
\delta=\frac{4}{9H^2} \frac{k^2}{a^2} {\cal R},
\end{equation}
where $a$ is the scale factor and we have assumed that the Universe is dominated
by radiation.
Thus, the density contrast evolves in proportion to $a^2$ on super-Hubble scales.
On this initial surface, we consider an overdense region whose size is $a_{\rm ini}R$ 
($R$ is the comoving size).
Then the density contrast of that region evolves as 
\begin{equation}
\delta (t)=\delta_{\rm ini} \left( \frac{t}{t_{\rm ini}} \right),
\end{equation}
on super-Hubble scales.
For both the PBHs and UCMHs, their formation criterion can be formulated as follows.
At the time of horizon crossing defined by
\begin{equation}
a_{\rm ini}R {\left( \frac{t_H}{t_{\rm ini}} \right)}^{1/2}=H^{-1}=2t_H,~~~
\longrightarrow ~~~t_H=\frac{a_{\rm ini}^2R^2}{4t_{\rm ini}},
\end{equation}
if the overdensity is larger than the critical density $\delta_{\rm th}$, 
such a region eventually collapses to form a bound object.
For the PBHs, we take $\delta_{\rm th}=0.5$ and $\delta_{\rm th}=10^{-3}$ for the UCMHs.
In reality, because the formation depends also on the density profile, we do not have
a unique value of $\delta_{\rm th}$ \cite{Shibata:1999zs,Polnarev:2006aa,Hidalgo:2008mv,1475-7516-2014-01-037}. But for simplicity, we do not consider such effects and
assume that collapse occurs when the overdensity exceeds the fixed value of $\delta_{\rm th}$.
Using the expression of $t_H$ given above,
the formation condition $\delta_H \ge \delta_{\rm th}$ can be written as
\begin{equation}
\delta_{\rm ini} \ge \Delta (R) \equiv \frac{4t_{\rm ini}^2}{a_{\rm ini}^2R^2}\delta_{\rm th}.
\end{equation}
According to the Press-Schechter formalism, the probability of having collapsed objects
in a range $(R,R+dR)$ is given by
\begin{equation}
-\frac{d\beta}{dR} dR=2\int_{\Delta (R)}^\infty \frac{\partial}{\partial R}P(\delta_{\rm ini},R)d\delta_{\rm ini}~dR.
\end{equation}
where $P(\delta_{\rm ini},R)$ is the probability density that the smoothed region 
of size $R$ has the density contrast $\delta_{\rm ini}$ at $t=t_{\rm ini}$.
Here the upper limit is assumed to be $\infty$, which is a good approximation both for the PBH case
and the UCMH case.
In our paper, we assume a Gaussian distribution of $P(\delta_{\rm ini},R)$, namely,
\begin{equation}
P(\delta_{\rm ini},R)=\frac{1}{\sqrt{2\pi} \sigma_R} \exp \left( -\frac{\delta_{\rm ini}^2}{2\sigma_R^2} \right).
\end{equation}
Note that $\sigma_R^2$ is the variance evaluated at $t=t_{\rm ini}$.
For the power spectrum given by Eq.~(\ref{ini-P}), we have
\begin{equation}
\sigma_R^2={\cal P}_{\delta,*} {\left( \frac{k_*}{a_{\rm ini}H_{\rm ini}} \right)}^4 e^{-k_*^2 R^2},
~~~~~{\cal P}_{\delta,*} =\frac{16}{81} {\cal P}_*. 
\end{equation}
For the Gaussian distribution function of $P(\delta_{\rm ini},R)$, we find
\begin{equation}
-\frac{d\beta}{dR} dR=-2 \frac{\partial \sigma_R}{\partial R} \frac{\Delta (R)}{\sigma_R} P(\Delta,R) dR.
\end{equation}
Or, in terms of the mass of the collapsed object $M$, this becomes
\begin{equation}
M\frac{d\beta}{dM}=-2 \frac{M}{R} \frac{dR}{dM} \frac{d\ln \sigma_R}{d\ln R} \Delta P(\Delta,R) dM. \label{mass-func}
\end{equation}
We now have the basic formula of the mass function.
In the following, we will apply the formula (\ref{mass-func}) first to
the PBH mass function and then to the UCMH mass function.

\subsubsection{Application to the PBH mass function}
In the radiation-dominated universe, 
if the density contrast of the overdense region is larger than $\delta_{\rm th}=0.5$
at the time of horizon crossing,
such a region undergoes gravitational collapse and becomes a PBH \cite{Carr:1974nx}.
The mass of the PBH is approximated by the horizon mass evaluated at the time of horizon crossing.
Thus, the mass of the PBH that formed from the overdense region with the comoving radius $R$
can be written as \footnote{
According to numerical simulations, a spherical overdense region
collapses to a BH when the radius of the region becomes equal to the Hubble radius $H^{-1}$.}
\begin{equation}
M=\frac{a_{\rm ini}^2 R^2}{4G t_{\rm ini}}.
\end{equation}
Note that
$M$ is independent of $t_{\rm ini}$ since $a_{\rm ini} \propto t_{\rm ini}^{1/2}$.
We then have
\begin{equation}
\frac{M}{R} \frac{dR}{dM}=\frac{1}{2},~~~~~\frac{d\ln \sigma_R}{d\ln R}=-k_*^2 R^2,
~~~~~\Delta=\frac{t_{\rm ini}}{GM}\delta_{\rm th}.
\end{equation}
Using these equations, we find
\begin{equation}
k_*^2 R^2=\frac{2GMk_*^2}{H_0 \sqrt{\Omega_{\rm rad}}},
\end{equation}
and then
\begin{equation}
\frac{\Delta}{\sigma_R}=\frac{H_0 \sqrt{\Omega_{\rm rad}}}{2{\cal P}_{\delta,*}^{1/2} k_*^2} 
\exp \left( \frac{GMk_*^2}{H_0 \sqrt{\Omega_{\rm rad}}} \right) \frac{\delta_{\rm th}}{GM}. \label{ds1}
\end{equation}
Finally, the mass function (\ref{mass-func}) can be written as
\begin{equation}
M \frac{d\beta}{dM} =\frac{2GMk_*^2}{H_0 \sqrt{\Omega_{\rm rad}}} \frac{1}{\sqrt{2\pi}} 
\frac{\Delta}{\sigma_R} \exp \left( -\frac{\Delta^2}{2\sigma_R^2} \right), \label{fin-mass-function}
\end{equation}
where Eq.~(\ref{ds1}) should be used for $\Delta/\sigma_R$ in the last equation.
As it should be, the mass function is independent of $t_{\rm ini}$.

For the case of our interest for which ${\cal P}_{\delta,*} \ll \delta_{\rm th}^2$ is always satisfied,
the mass function (\ref{fin-mass-function}) has a sharp peak
and thus the mass at the peak gives the typical PBH mass.
The peak mass $M_{\rm peak}$ can be evaluated analytically by the condition that 
it minimizes the ratio $\Delta/\sigma_{\rm{R}}$,
yielding
\begin{equation}
\frac{GM_{\rm peak} k_*^2}{H_0 \sqrt{\Omega_{\rm rad}}}=1, ~~~~~\to ~~~~~M_{\rm peak}=2 \times 10^4~M_\odot 
~{\left(\frac{k_*}{5 \times 10^4~{\rm Mpc}^{-1}} \right)}^{-2}. \label{pbh-peak-mass}
\end{equation}
The total fraction of the PBHs is given by the integral of $\frac{d\beta}{dM}$ over $M$.
The mass function around the peak mass can be approximately written as
\begin{equation}
M \frac{d\beta}{dM} \approx \fr{{\rm e}\delta_{\rm{th}}}{\sqrt{2\pi {\cal P}_{\delta,*}}}
\exp\left(-\frac{{\rm e}^2\delta_{\rm{th}}^2}{8{\cal P}_{\delta,*}}\right)
\exp \left( - \frac{{\rm e}^2 \delta_{\rm th}^2}{8{\cal P}_{\delta,*} M_{\rm peak}^2} {(M-M_{\rm peak})}^2 \right),
\end{equation}
where ${\rm e}$ is the Napier's constant.
Using this equation, $\beta$ is written as
\begin{equation}
\beta = \int ~\frac{d\beta}{dM}dM=2 \exp \left( -\frac{{\rm e}^2 \delta_{\rm th}^2}{8 {\cal P}_{\delta,*}} \right).
\end{equation}
Given $\beta$ and $\delta_{\rm th}$, we can determine ${\cal P}_{\delta,*}$
as a solution of the above equation.
Because of different dilution rates of radiation and PBHs due to the cosmic expansion,
the current abundance $\Omega_{\rm PBH}$ of PBHs is enhanced by the duration of 
the radiation dominated epoch measured since the PBH formation time.
Thus, we have
\be
\beta=
\fr{\Omega_\m{PBH}}{\Omega_\m{m}}\fr{1+z_\m{eq}}{1+z_\m{PBH}}, \label{omega-pbh}
\ee
where $z_\m{PBH}$ is the cosmological redshift when the PBHs are produced.
Now, let us estimate $\Omega_\m{PBH}$ for some interesting cases.
If the SMBHs residing in present galaxies are PBHs,
assuming a seed PBH of mass $M_{\rm PBH}$ is contained in each galaxy,
$\Omega_\m{PBH}$ is estimated as
\begin{equation}
\Omega_\m{PBH} \simeq \Omega_m \frac{M_{\rm PBH}}{M_{\rm gal}}
=3 \times 10^{-9}~\left( \frac{M_{\rm PBH}}{10^4~M_\odot} \right) 
{\left( \frac{M_{\rm gal}}{10^{12}~M_\odot} \right)}^{-1}. \label{omega-pbh-gal}
\end{equation}
Since the PBH mass is equal to the horizon mass at the formation time,
we have
\begin{equation}
M_{\rm PBH}=\frac{4\pi}{3} \rho (z_{\rm PBH}) H^{-3}(z_{\rm PBH})
\simeq 5\times 10^{17} M_\odot ~{\left( \frac{1+z_{\rm PBH}}{1+z_{\rm eq}}\right)}^{-2}. \label{mass-pbh}
\end{equation}
Using Eqs.~(\ref{omega-pbh-gal}) and (\ref{mass-pbh}), we find
\begin{equation}
\beta \simeq 10^{-15} {\left( \frac{M_{\rm PBH}}{10^4~M_\odot} \right)}^{3/2} 
{\left( \frac{M_{\rm gal}}{10^{12}~M_\odot} \right)}^{-1},
\end{equation}
corresponding to ${\cal P}_{\delta,*}=7\times 10^{-3}$
for $\beta=10^{-15},~\delta_{\rm th}=0.5$.
For the case in which the SMBHs observed at high redshifts $(z=6 \sim 7)$ are PBHs,
using the comoving number density of the SMBHs $n_{\rm BH} \simeq 1~{\rm Gpc}^{-3}$,
we have 
\be
\Omega_\m{PBH} = \frac{M_{\rm PBH} n_{\rm BH}}{\rho_c} = 8\times 10^{-17}~ 
\left(\frac{M_{\rm PBH}}{10^4~M_\odot} \right),~~~\longrightarrow~~~ 
\beta=4\times 10^{-23} {\left( \frac{M_{\rm PBH}}{10^4~M_\odot} \right)}^{3/2},
\ee
where $\rho_c$ is the critical density.
Setting $\beta=4\times 10^{-23},~\delta_{\rm th}=0.5$ leads to ${\cal P}_{\delta,*}=4\times 10^{-3}$.

From these examples, we find that ${\cal P}_{\delta,*}$ varies only by an ${\cal O}(1)$ factor even if 
we vary $\beta$ by several orders of magnitude,
which is simply because ${\cal P}_{\delta,*}$ depends on $\beta$ only logarithmically.
Thus, uncertainty in the PBH abundance does not affect our estimate of ${\cal P}_{\delta,*}$
so much. 
Furthermore, the abundance and mass of the UCMHs become insensitive to the value of
${\cal P}_{\delta,*}$ as long as it is much bigger than $\delta_{\rm th}^2$ for the UCMH.
Since $\delta_{\rm th} \simeq 10^{-3}$ for the UCMH, ${\cal P}_{\delta,*} \gg \delta_{\rm th}^2$
is always satisfied in the PBH scenario.
To conclude, uncertainties such as the PBH abundance and $\delta_{\rm th}$ for the PBH
have little impact on our constraint on the dark matter annihilation.

Lastly, let us consider the minimum PBH mass for which our constraints
based on the copious UCMH formation can be applied.
Density perturbations whose length scale is shorter than the free streaming scale
can not grow to form UCMHs.
The comoving wave number for the free streaming scale is given by \cite{Green:2005fa}
\begin{equation}
k_{\rm fs} \simeq 3~{\rm pc}^{-1}~{\left( \frac{m_\chi}{1~{\rm TeV}} \right)}^{1/2}
{\left( \frac{T_{\rm kd}}{10~{\rm MeV}} \right)}^{1/2},
\end{equation}
where $m_\chi$ is the WIMP mass and $T_{\rm kd}$ is the temperature
below which WIMPs are kinematically decoupled.
Then, from Eq.~(\ref{pbh-peak-mass}), the PBH mass corresponding to $k_{\rm fs}$ is given by
\begin{equation}
M_{\rm PBH}=6~M_\odot {\left( \frac{m_\chi}{1~{\rm TeV}} \right)}^{-1}
{\left( \frac{T_{\rm kd}}{10~{\rm MeV}} \right)}^{-1}.
\end{equation}
For definiteness, we take $1~M_\odot$ as the minimum PBH mass when
we give plots of our results as a function of $M_{\rm PBH}$.

\subsubsection{Application to the UCMH mass function}
As mentioned in the Introduction, 
if the density contrast of an overdense region is as large as ${\cal O}(10^{-3})$
at the horizon crossing time,
such a region undergoes gravitational collapse at around the time of the
recombination and forms a gravitationally bound object (UCMH) \cite{Ricotti:2009bs}.
Assuming that only the DM inside the overdense region collapses to a UCMH, 
the UCMH mass $M$ is given by
\begin{equation}
M=\frac{1}{2G}{\left(H_0 \sqrt{\Omega_{\rm rad}}\right)}^2 (1+z_{\rm eq}) R^3. \label{ucmh-mass}
\end{equation}
As before, $R$ is the comoving radius of the overdense region.
Then, we have
\begin{equation}
\frac{M}{R} \frac{dR}{dM}=\frac{1}{3},
\end{equation}
and
\begin{equation}
k_*^2 R^2=k_*^2 {\left( \frac{2GM}{{\left(H_0 \sqrt{\Omega_{\rm rad}}\right)}^2(1+z_{\rm eq})} \right)}^{2/3},
\end{equation}
as well as
\begin{equation}
\frac{\Delta}{\sigma_R}=\frac{\delta_{\rm th}}{{\cal P}_{\delta,*}^{1/2} k_*^2 R^2} 
\exp \left( \frac{k_*^2R^2}{2} \right).
\end{equation}
Using these relations, the mass function can be written as
\begin{equation}
M \frac{d\beta}{dM}= \frac{2}{3} k_*^2R^2 \frac{1}{\sqrt{2\pi}} \frac{\Delta}{\sigma_R} \exp \left( -\frac{\Delta^2}{2\sigma_R^2} \right).
\end{equation}
In this case too, the eventual result is independent of $t_{\rm ini}$.

Contrary to the PBH case, the peak mass is 
larger than the total dark matter mass contained in the 
Hubble horizon evaluated at the horizon crossing time of the scale $R$.
This is simply because the overdense region corresponding to the peak mass is larger than
the scale $R$.

The upper limits on the annihilation cross sections we will derive later using the mass function obtained here 
are conservative for the following reason.
In our case, the density contrast on the comoving scale $R$ is as large as 
$\sqrt{{\cal P}_{\delta,*}}=0.06$ and the density contrast of $10^{-3}$ is achieved on
a larger smoothing scale.
In reality, smaller and denser UCMHs formed from the perturbation peaks of scale $R$ should exist as well.
In other words, a UCMH of the peak mass is expected to contain a population
of smaller and denser UCMHs.
However, the mass function computed in the present formalism neglects those smaller UCMHs and
treat the larger UCMHs as structureless objects.
Since annihilation occurs more efficiently in the smaller and denser UCMHs, 
the neglect of the lumpy structures results in an underestimate of 
the cosmic ray flux coming from the UCMHs, hence a conservative estimation. 

The typical UCMH mass, given by the peak mass,
is determined by the larger solution $R_{\rm{c}}$ of 
$\Delta/\sigma_{R_{\rm{c}}}=1$,
namely,
\begin{equation}
\frac{\delta_{\rm th}}{{\cal P}_*^{1/2} k_*^2 R_{\rm{c}}^2} \exp \left( \frac{k_*^2R_{\rm{c}}^2}{2} \right)=1.
\end{equation}
Defining $k_*^2R_{\rm{c}}^2/2\equiv x_{\rm{c}}$ and 
$C\equiv \delta_{\rm{th}}/2{\cal P}_{\delta,*}^{1/2}=8\times 10^{-3}$ with 
$\delta_{\rm{th}}=10^{-3},~{\cal P}_{\delta,*}=4\times 10^{-3}$,
$x_{\rm{c}}$ is approximately given by
\begin{equation}
x_{\rm{c}}\simeq \log\left\{\frac{1}{C}\log\left(\frac{-\log C}{C}\right)\right\}=6.7.
\ee
Note that the smallness of the ratio $C$ implies that a sizable fraction ($\sim {\cal O}(1)$) 
of dark matter is contained in UCMHs. 
Now, since $k_*$ is related to $M_{\rm BH}$ by Eq.~(\ref{pbh-peak-mass})
and $R$  to $M_{\rm UCMH}$ by Eq.~(\ref{ucmh-mass}), we can
obtain the relationship between $M_{\rm UCMH}$ and $M_{\rm BH}$ by using the definition of $x_{\rm{c}}$ as
\begin{equation}
M_{\rm UCMH}=(1+z_{\rm eq}) x_c^{3/2} \sqrt{2GH_0 \sqrt{\Omega_{\rm rad}}} ~M_{\rm BH}^{3/2}
=3\times 10^{-2}~M_\odot~{\left( \frac{M_{\rm BH}}{10^4~M_\odot}\right)}^{3/2}.
\end{equation}

We can also estimate the radius of the UCMH, 
which is approximately half the radius of the overdense region 
at the moment of the turn around.
From the definition of $x_{\rm{c}}$, 
\begin{equation}
R_{\rm{UCMH}}\simeq \frac{a_{\rm turn}R_{\rm turn}}{2}=\frac{1}{1+z_{\rm turn}}
\sqrt{\frac{x_c GM_{\rm BH}}{2H_0 \sqrt{\Omega_{\rm rad}}}}
=2 \times 10^{11}~{\rm km} \left( \frac{1+z_{\rm eq}}{1+z_{\rm turn}} \right)
{\left( \frac{M_{\rm BH}}{10^4~M_\odot}\right)}^{1/2}.
\end{equation}
In this paper $\zt$ is set to 1000 following previous work \cite{Ricotti:2009bs}. 
It should be noted that subsequent interactions and mergers among
UCMHs and baryonic matter may change this picture (though the analysis of \cite{Berezinsky:2007qu}
suggests that these effects are not significant). 
An analysis of such effects is completely beyond the scope of this work.

\subsection{Properties of UCMHs}
As for the density profile of the UCMHs,
we adopt the following form
\begin{equation}
\rho_\chi (r)=\rho_0 {\left( 1+\frac{r}{r_c} \right)}^{-\alpha}, \label{ucmh-profile}
\end{equation}
where $r_c$ is the core radius which is to be determined shortly.
The secondary infall theory based on the spherical collapse model predicts $\alpha=9/4$
\cite{Fillmore:1984wk, Bertschinger:1985pd}
and this value has been used in the literature 
\cite{Ricotti:2009bs,Josan:2010vn,Bringmann:2011ut,Li:2012qha,Yang:2013dsa},
whereas $\alpha$ was set to $\sim$1.8 in \cite{Berezinsky:2007qu}. 
Given this situation, we leave $\alpha$ unspecified in a range $1.5<\al <3$
and study the dependence of the eventual results on $\alpha$. 
Then, the mass of the dark matter inside the radius $r(>r_c)$ is
\begin{equation}
M(r) = \frac{4\pi \rho_0 r^3}{3-\alpha} {\left( \frac{r_c}{r} \right)}^\alpha.
\end{equation}
The condition $M_{\rm UCMH}=M(R_h)$ gives $\rho_0$ in terms of other quantities as
\begin{equation}
\rho_0 =\frac{3-\alpha}{4\pi R_h^3} M_{\rm UCMH} {\left( \frac{r_c}{R_h}\right)}^{-\alpha}.
\end{equation}
There are two different processes that contribute to smoothing of the 
density profile near the center and determine the value of $r_c$ \cite{Bringmann:2011ut}.
The first one is concerned with actual dark matter particles having velocity dispersion.
If it were not for velocity dispersion and the collapse were purely radial,
then the mass density would be singular at the center. 
In reality, the radial infall breaks down at some radius $r_{\rm Kep}$, where the 
angular velocity becomes equal to the Kepler velocity:
\begin{equation}
v_{\rm Kep} (r_{\rm Kep})=\sqrt{\frac{GM(r_{\rm Kep})}{r_{\rm Kep}}}.
\end{equation}
We use the velocity dispersion $\sigma_\chi$ at the inception of the collapse derived in 
\cite{Ricotti:2007au,Ricotti:2009bs}
\begin{equation}
\sigma_\chi (\zt)=\sigma_{\chi,0} {\left( \frac{1000}{1+\zt} \right)}^{1/2} {\left( \frac{M_{\rm UCMH}}{M_\odot} \right)}^{0.28},
\end{equation}
where $\sigma_{\chi,0} =0.14~{\rm m/s}$.
Using the conservation of the angular momentum,
the radius $r$ at which the radial infall breaks down is written as
\begin{eqnarray}
\fr{r_{\rm Kep}}{R_h}&=&{\left( \sigma_\chi^2 \frac{R_h}{GM_{\rm UCMH}} \right)}^{\frac{1}{4-\alpha}} \nonumber \\
&=&\left[
7\times 10^{-8}\left( \frac{1+z_{\rm eq}}{1+\zt}\right)^2 
\left( \frac{M_{\rm PBH}}{10^4~M_\odot} \right)^{-0.16}
\right]^{\fr{1}{4-\al}}. 
\end{eqnarray}
When $r_\m{Kep}$ is sufficiently large to satisfy the following inequality, 
the central region of a UCMH is smoothed due to the angular momentum without annihilation being effective: 
\be
\frac{3-\alpha}{4\pi R_h^3} M_\m{UCMH} {\left( \frac{r_\m{Kep}}{R_h}\right)}^{-\alpha}
<\rho_\m{anni}\simeq\frac{m_\chi}{\langle \sigma v \rangle t_0},
\ee
where $t_0 \approx 13.7~{\rm Gyr}$ is the age of the Universe.
This inequality can be rewritten as the condition for the mass of PBHs as follows:
\be
\fr{M_\m{PBH}}{10^4M_\odot}
<2\times 10^{-45}
\left(
\fr{1+z_\m{eq}}{1+\zt}
\right)^{12.5}
\left[
\fr{8\times 10^5}{3-\al}
\left(
\fr{1+z_\m{eq}}{1+\zt}
\right)^3
\left(
\fr{m_\chi}{1\m{TeV}}
\right)
\left(
\fr{\langle\si v\rangle}{\langle\si v\rangle_{\mathrm{can}}}
\right)^{-1}
\right]^{\fr{4-\al}{0.16\al}}\label{keporanni},
\ee
where $\langle\si v\rangle_{\mathrm{can}}=3\times10^{-26}\m{cm}^3/\m{s}$ is the canonical value.
In contrast, when this inequality does not hold, 
the central region of a UCMH is smoothed due to annihilation. 
In this case, the central density $\rho_0$ becomes equal to $\rho_\m{anni}$:
\be
\frac{3-\alpha}{4\pi R_h^3}M_\m{UCMH}\left(\frac{r_\m{c}}{R_h}\right)^{-\alpha}
=\rho_\m{anni}.
\ee
Denoting the core radius in this case as $r_\m{c}=r_\m{anni}$, 
this equation yields the following expression for $r_\m{anni}$:
\begin{eqnarray}
\fr{r_{\rm anni}}{R_h} 
&=&\left[
 \frac{\langle \sigma v \rangle t_0}{m_\chi} \frac{(3-\alpha) M_{\rm UCMH}}{4\pi R_h^3} 
 \right]^{\frac{1}{\alpha}} \nonumber \\
&=&\left[
\fr{8\times 10^5}{3-\al}
\left(
\fr{1+z_\m{eq}}{1+\zt}
\right)^3
\left(
\fr{m_\chi}{1\m{TeV}}
\right)
\left(
\fr{\langle\si v\rangle}{\langle\si v\rangle_{\mathrm{can}}}
\right)^{-1}
\right]^{-1/\al}\label{keporanni}.
\end{eqnarray}
Fig.~\ref{ns_figures} shows which mechanism dominates and determines the core radius $r_c$,
which is the larger of $r_{\rm Kep}$ or $r_{\rm anni}$.
\begin{figure}[t]
\begin{center}
\includegraphics[width=13cm,keepaspectratio,clip]{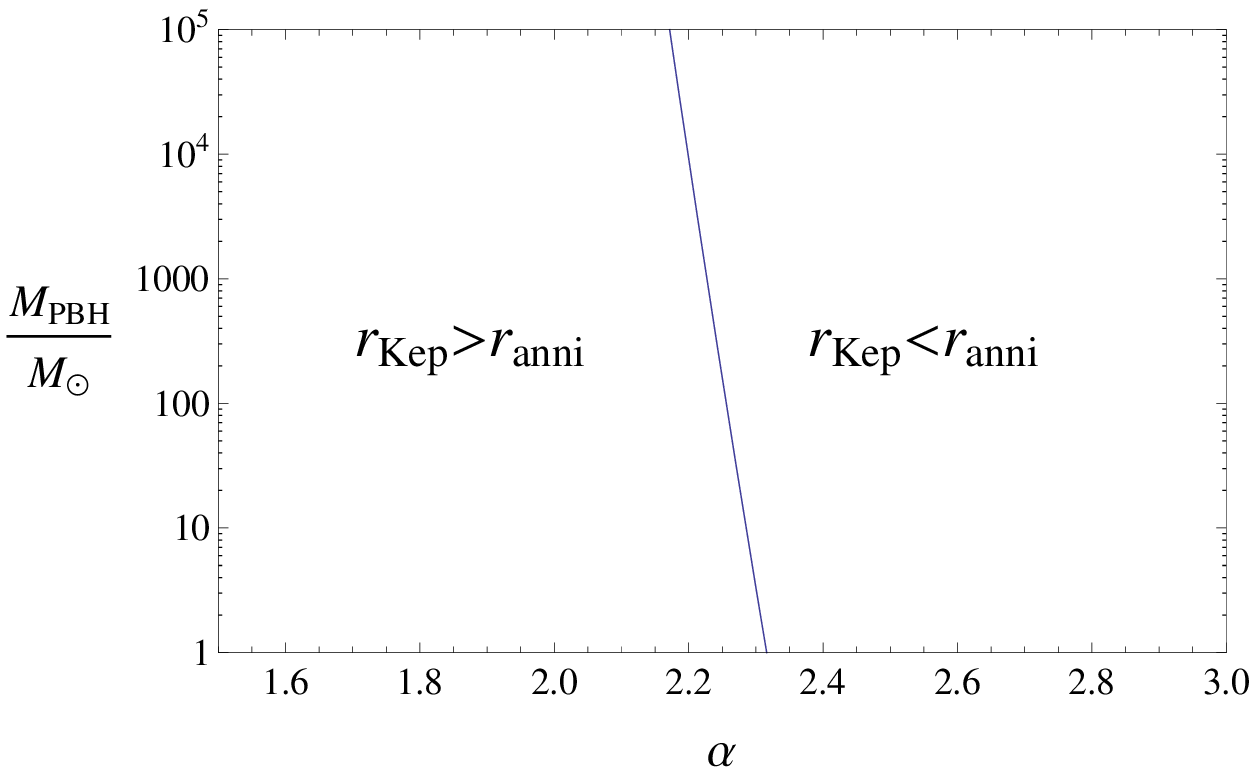}
\end{center}
\caption{
The curve of $\MP$ determined by (\ref{keporanni}) 
with $m_\chi=1$TeV and $\langle \sigma v\rangle=\langle\si v\rangle_{\mathrm{can}}$, below (above) which $\rK>\ra\quad (\rK<\ra)$, when the profile of UCMHs is smoothed 
due to Kepler motion (annihilation). 
}
\label{ns_figures}
\end{figure}
\subsection{Observational constraints on the cross sections from $\gamma$-rays}
Before comparing the expected cosmic-ray flux from the UCMHs with the 
observational data,
let us first consider whether $\gamma -$rays from the UCMHs in the Milky 
Way are observed as those from point sources or as diffuse $\gamma -$ray background. 
Assuming about half of the entire dark matter is contained in UCMHs,  
the number of UCMHs in our galaxy, whose mass is denoted by $M_\m{MW}$, is 
estimated by 
\be
N_\m{UCMH}\sim\fr{M_\m{MW}}{2\MU}.
\ee
Angular resolution $\theta^{\circ}$ of Fermi-LAT depends on the photon energy; 
for the energy interval covered by the Fermi-LAT, $100~{\rm MeV} \sim 600~{\rm GeV}$, 
$\theta$ is a monotonically decreasing function of the photon energy with 
$\theta=5^\circ$ at $100~{\rm MeV}$ and $\theta=0.1^\circ$ at $100~{\rm GeV}$ 
\footnote{
http://www.slac.stanford.edu/exp/glast/groups/canda/lat\_Performance.htm
}. 
The number of UCMHs contributing to $\gamma -$rays observed within 
the solid angle corresponding to $\theta$ is given by
\be
N_\theta=N_\m{UCMH}\times\fr{A_\theta}{4\pi}
,
\ee
where $A_\theta$ denotes the area of a circular region on a unit sphere subtended by $\theta/2$:
\be
A_\theta=\pi\times\left(\fr{1}{2}\times\fr{2\pi}{360}\times\theta\right)^2.
\ee
If this quantity $N_\theta$ is sufficiently larger than unity, 
UCMHs should be contributing to the diffuse emission, 
otherwise UCMHs are observed as point sources. 
Since the number of the UCMHs is dependent on their mass,
whether the cosmic-rays from UCMHs can be seen as diffuse or
separate points also depends on the UCMH mass.
As we will see later, the most stringent bounds on DM annihilation modes 
are set by the photon flux measured at energy around $100~{\rm GeV}$ for 
DM particle mass $1~{\rm TeV}$, which we assume here as a typical value. 
Then, using $\theta=0.1^\circ$ for the photon energy $\sim 100~{\rm GeV}$,
the critical mass dividing the two regimes (whether diffuse or point source) 
is given by 
\be
M_\m{PBH}=10^9\MS\left(\fr{\theta}{0.1^{\circ}}\right)^{4/3}
\left(\fr{M_\m{MW}}{10^{12}\MS}\right)^{2/3}.
\ee
For PBHs less massive than this value, UCMHs are so dense that emission 
from them can be observed as diffuse emission. 
Since in this section we focus on PBHs whose mass is at most $\sim 5\times 10^4\MS$, 
$\gamma$-rays from UCMHs can be regarded as diffuse and so 
later are compared with what is known as extra-galactic $\gamma$-ray emission.

Let us now consider constraints imposed by the Fermi-LAT on the $\gamma$-ray
flux coming from the UCMHs in the Milky Way
\footnote{
The $\gamma$-ray flux originating from extra-galactic UCMHs has been estimated to be 
at most comparable to that from UCMHs in the Milky Way and therefore is not considered
here.
}.
We use the observed $\gamma$-ray flux as the upper bound on the $\gamma$-ray
coming from the UCMHs and translate it to upper bounds on the annihilation
cross sections to the SM particles \footnote{
In a similar manner to the analysis done in this paper, 
it is straightforward to place limits on the decay rate of DM particles if 
they decay into SM particles.
However, the $\gamma -$ray flux in this case does not differ significantly from the
case of no UCMHs, contrary to the annihilation case in which the strong flux comes
from the core of UCMHs. Thus, the DM decay case is not relevant to the test of the PBH scenario
and we do not consider this case in this paper.
}.
As the density profile of the dark matter in the Milky Way, we assume the Navarro-Frenk-White (NFW) profile 
\footnote
{
Note that what matters to the evaluation here is the dark matter profile of only the integrated part 
in eq.(\ref{diffuse}) and the dark matter profile in the center of our galaxy, the highly controversial part, is irrelevant. 
}
given by\cite{Battaglia:2005rj}
\begin{equation}
\rho_{\rm MW}(r)=\frac{\delta_c \rho_c^0}{(r/r_s) {(1+r/r_s)}^2},
\end{equation}
where $\rho_c^0$ is the critical density and $r_s$ is the scale length.
\begin{figure}[t]
\begin{center}
\includegraphics[width=9.7cm,keepaspectratio,clip]{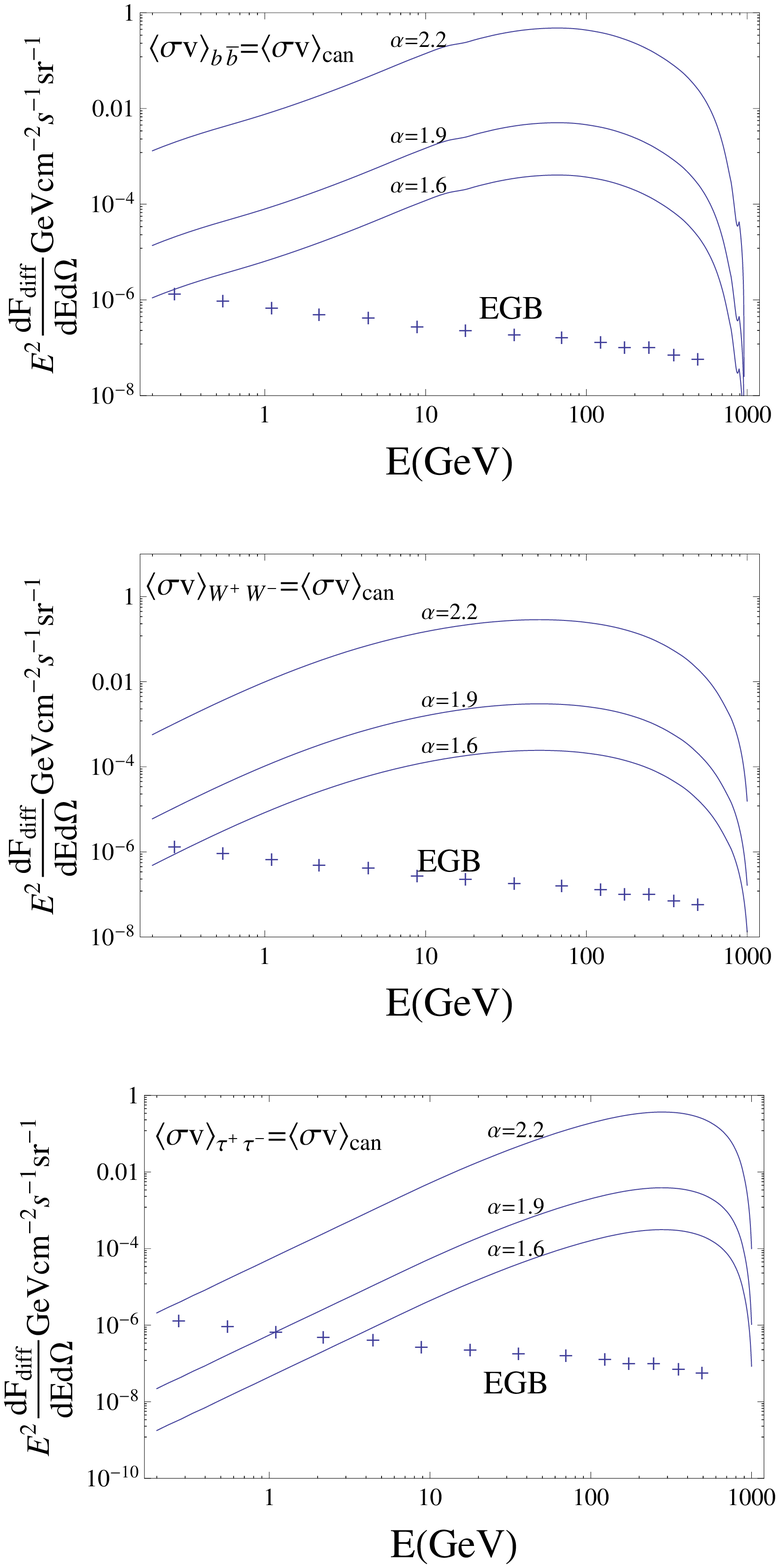}
\end{center}
\caption{
The energy flux of $\gamma -$rays from UCMHs calculated by Eq.~(\ref{diffuse}) 
with each panel obtained assuming only one mode is dominant and the annihilation cross section of this dominating mode is the canonical value: $\langle \sigma v\rangle_k=\langle \sigma v\rangle_{\mathrm{can}}=3\times 10^{-26}\rm{cm}^3/s$. 
Here, $m_\chi=1$TeV, $\MP=10^4\MS$ and $z_{\rm{turn}}=1000$ are assumed and these values are also used in other plots.
For comparison, the extra-galactic $\gamma$-ray background (EGB) inferred by the observation of Fermi-LAT is also presented by crosses. 
}
\label{flux}
\end{figure}
\begin{figure}[t]
\begin{center}
\includegraphics[width=10.6cm,keepaspectratio,clip]{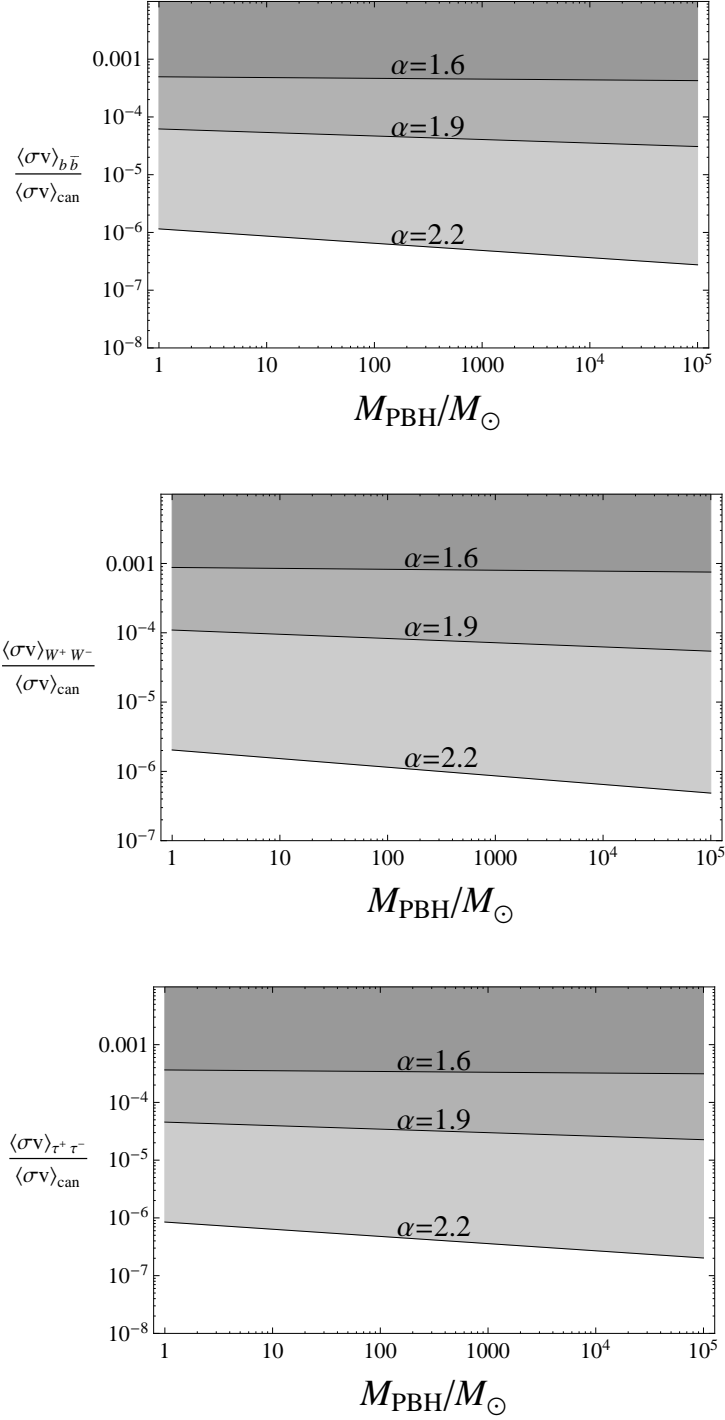}
\end{center}
\caption{
The constraint on the cross section of each 
mode, normalized 
by the canonical value $\langle \sigma v \rangle_{\rm{can}} =3\times 10^{-26} {\rm cm}^3/{\rm s}$, as a function of $\MP$. 
For each value of $\alpha$, the shaded region is excluded. 
}
\label{gammaconstmpbh}
\end{figure}
\begin{figure}[t]
\begin{center}
\includegraphics[width=11cm,keepaspectratio,clip]{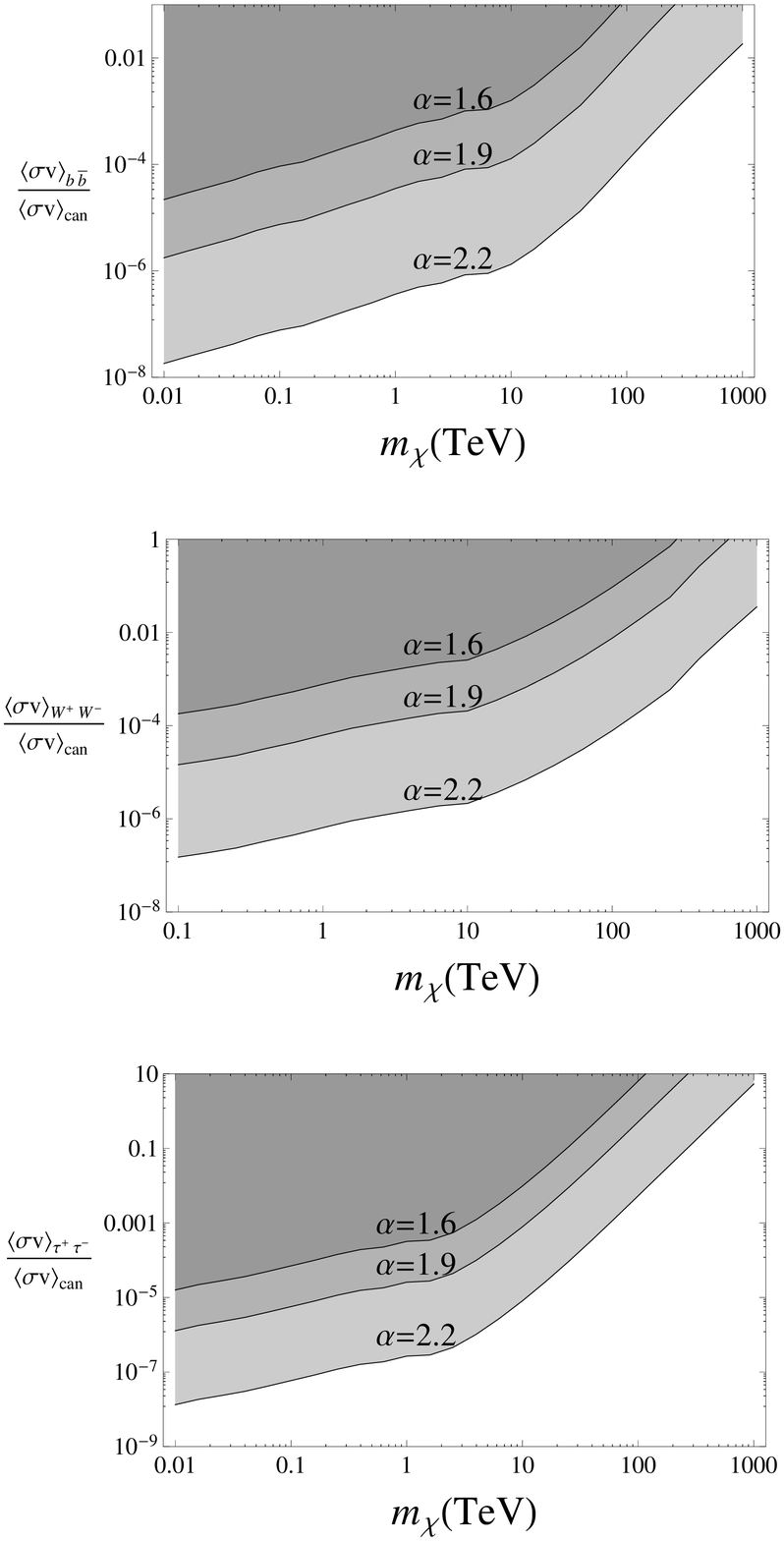}
\end{center}
\caption{
The dependence of the constraints of the cross sections on the dark matter mass.
}
\label{gammaconstmdm}
\end{figure}
Since we do not know {\it a priori} properties of the DM particles,  
we consider three annihilation modes from DM to SM particles ($\chi \chi \to b{\bar b}$, 
$\chi \chi \to W^+ W^-$ and $\chi \chi \to \tau^+ \tau^-$), individually. 
The $\gamma$-ray flux $\Phi_{\gamma,k}$ created by an annihilation mode labeled by $k$
inside a UCMH located at a distance $d$ from the Earth is measured as:
\begin{equation}
4\pi d^2 \Phi_{\gamma,k}=\int_{\rm UCMH} d^3x ~N_{\gamma,k} \frac{\rho_\chi^2 \langle \sigma v \rangle_k}{2m_\chi^2}, \label{gamma-flux}
\end{equation}
where $\langle \sigma v \rangle_k$ represents the cross section of the mode $k$ and 
$N_{\gamma,k}$ is the average number of created photons due to one annihilation process $k$.
In what follows, we consider each annihilation mode separately and place a limit for each mode 
independently.
Since combining all the annihilation modes only increases the $\gamma -$ray
flux, our results provide conservative upper bounds on the cross section of each mode. 
Then, the energy flux coming from UCMHs at unit solid angle is calculated as follows:
\begin{eqnarray}
E^2\fr{dF_\m{diff}}{dEd\Om}
&=&\fr{1}{2\MU}\int^{d_\m{MW}}_{d_\m{E}}\rho_\m{MW}(d')E^2\fr{d\Phi(d')}{dE}d'^2\m{d}d'\nonumber\\
&=&\fr{1}{2\MU}E^2\fr{dN_{\gamma,k}}{dE}\fr{\langle\sigma v\rangle_k}{2m_\chi^2}\int^{R_{\m{UCMH}}}_{0}r^2\rho_{\chi}^2(r)dr\int^{d_{\m{MW}}}_{d_\m{E}}\rho_{\m{MW}}(d')\m{d}d'\nonumber\\
&=&1.2\times 10^{-6}\m{s^{-1}cm^{-2}sr^{-1}}\fr{\alpha(3-\alpha)^2}{2\alpha-3}E^2\fr{dN_{\gamma,k}}{dE}\left(\fr{m_\chi}{1\m{TeV}}\right)^{-2}\nonumber\\
&&\quad\quad\quad\quad\quad\times\left(\fr{\langle\sigma v\rangle_k}{\langle\si v\rangle_{\mathrm{can}}}\right)
\left(\fr{1+z_\m{eq}}{1+\zt}\right)^{-3}\left(\fr{r_\m{c}}{R_\m{h}}\right)^{3-2\alpha},
\label{diffuse}
\end{eqnarray}
where, $d_\m{MW}\sim 300\m{kpc}$ is the radius of our galaxy and 
$d_{E}\sim 8\m{kpc}$ is the distance between the center of our galaxy and Earth. 
The NFW profile has been integrated from the location of the Earth to 
the closest edge of our galaxy along the galactic plane.
The NFW profile has been approximated by $\propto r^{-1}$ inside the scale length 
and by $\propto r^{-3}$ outside. 
The quantity $dN_{\gamma,k}/dE$ is the number of the emitted photons in one annihilation 
per unit energy, 
which is calculated using PYTHIA\cite{Sjostrand:2006za}. 
Shown in 
Fig.~\ref{flux} is the calculated energy flux of $\gamma -$rays 
by using the formula (\ref{diffuse}) for three modes $\chi \chi \to b{\bar b},\; \chi \chi \to W^+ W^-$ and $\chi \chi \to \tau^+ \tau^-$, with each cross section set to the canonical value. 
For comparison, we also plot two sigma upper bounds of extra-galactic $\gamma$-ray background (EGB)\cite{Abdo:2010nz} obtained by the
Fermi-LAT
\footnote{
In Fig.~\ref{flux}, we also used data points of $100$GeV$<E$, which were shown in the presentation by 
M. Ackermann on behalf of the Fermi collaboration: http://agenda.albanova.se/conferenceDisplay.py?confId=2600.
}.
Clearly, the results show that setting the cross sections to the canonical value
is totally inconsistent with the Fermi-LAT data, {\it i.~e.}
the $\gamma -$ray flux in such cases by far exceeds the measured values.

The requirement that the estimated flux coming from UCMHs 
using the equation above should not exceed EGB 
is translated into the upper limits on the cross section $\langle \sigma v \rangle_k$, 
which is shown in Fig~\ref{gammaconstmpbh} and Fig~\ref{gammaconstmdm} as a function of $\MP$ and of $m_\chi$, respectively. 
There are several things worth mentioning.
First, interestingly, the dependence of the constraints on $M_{\rm PBH}$ is very weak. 
This is because the number density of UCMHs is inversely proportional to $\MU$, whereas 
the volume integral of the squared energy density of the dark matter inside a UCMH is almost proportional to the volume of the core ($\sim r_c^3$), 
which is almost proportional to $R_h^3\propto\MU$, 
hence the weak dependence on $\MU$, namely on $\MP$. 
Thus, the theoretical uncertainty of the increase of the PBH mass after the PBH formation
does not affect our final results so much.
Secondly, the constraints become tighter for larger values of $\alpha$. 
This may result from more concentrated density profile of the UCMH,
hence stronger $\gamma -$ray signals, for larger $\alpha$.
Thirdly, the constraints are weaker for larger values of $m_\chi$. 
When $m_\chi$ is increased, 1. the maximum value of the energy flux decreases because of the lower dark matter density and 
2. the position of the peak of the energy flux is shifted to higher energy. 
The first effect acts to make the constraints weaker,  
whereas the second effect acts in the opposite direction
since the EGB flux is lower for higher energy, but only when $m_\chi\lesssim 5$TeV because of the unavailability of 
the EGB data above $\sim 500$GeV.
It turns out that the first effect dominates the second, which explains the weaker constraints for lager $m_\chi$. 
Note that this tendency becomes even more noticeable for $m_\chi>10$TeV, which is because 
the unavailability of EGB data above $\sim 500$GeV invalidates the second effect. 
Lastly, 
the upper bounds for all the modes are mostly a lot smaller than the canonical value. 
This puts severe restriction on many WIMP scenarios in extended particle physics models where the
cross sections to standard model particles such as those considered in this paper are typically
around the canonical value unless some fine-tuning among model parameters is imposed.
In other words, our results indicate that if DM particles turn out to be WIMPs having
typical strength of interactions with SM particles, the PBH scenario as the origin
of SMBHs observed at high redshifts is strongly disfavored.

\subsection{Observational constraints on the cross sections from neutrino flux}
It is also worthwhile to consider constraints on the cross sections
obtained from neutrino flux and compare two kinds of constraints, 
obtained from $\gamma-$rays and neutrinos
\footnote{
Related to this section is \cite{Yang:2013dsa}, in which constraints on power spectrum of primordial perturbation were obtained by neutrino flux from UCMHs.
}.
Neutrino flux from UCMHs can be calculated in the same way as $\gamma -$rays discussed so far, except for 
the necessity to consider neutrino oscillation, whose effects can be incorporated using the following 
formula\cite{Lee:2012pz}:
\be
F_{\nu_\mu}
=0.24F_{\nu_e}^0
+0.40F_{\nu_\mu}^0
+0.35F_{\nu_\tau}^0,
\ee
where the subscript 0 indicates flux emitted from UCMHs 
and $F_{\nu_\mu}$ is the flux 
of $\nu_\mu$ at the Earth.
Neutrino flux thus calculated is shown in Fig. \ref{nuflux}, 
with that of two sigma upper bounds of atmospheric neutrinos, obtained by Frejus and IceCube
\footnote{
https://indico.triumf.ca/getFile.py/access?contribId=9\&sessionId=8\&resId=0\&materialId=slides\&confId=1756
}
, shown together for comparison.

The neutrino flux from UCMHs should not exceed that of atmospheric neutrinos, 
from which constraints on the cross sections are obtained as shown in Fig. \ref{nuconstmpbh} and \ref{numdm}, 
as a function of $\MP$  and of $m_\chi$, respectively. 
As in the case of photons, 
when $m_\chi$ is increased, 1. the maximum value of the energy flux decreases because of the lower dark matter density and 
2. the position of the peak of the energy flux is shifted to higher energy, 
with the first effect acting to make the constraints weaker,  
the second effect acting in the opposite direction. 
Though in Fig. \ref{numdm} the limit lines are zigzag, reflecting fluctuations of data points, 
the second effect dominates the first, and the constraints are overall tighter for larger $m_\chi$.  

\begin{figure}[t]
\begin{center}
\includegraphics[width=10.4cm,keepaspectratio,clip]{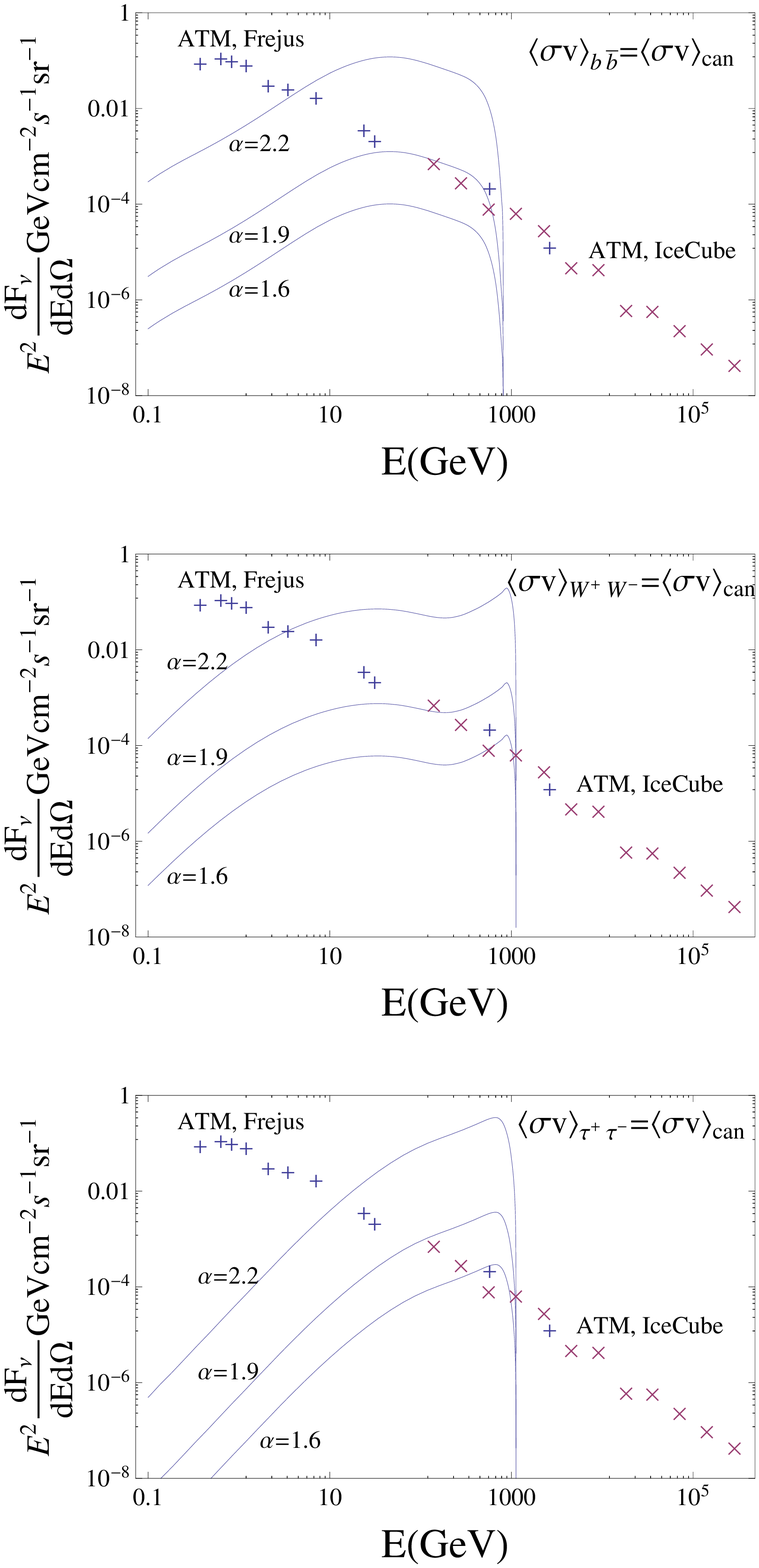}
\end{center}
\caption{
The energy flux of neutrinos from UCMHs calculated by Eq.~(\ref{diffuse}). 
The same parameter values as Fig. \ref{flux} are used. 
For comparison, the energy flux of the atmospheric neutrinos, obtained by Frejus and IceCube, is also presented. 
}
\label{nuflux}
\end{figure}
\begin{figure}[t]
\begin{center}
\includegraphics[width=10.7cm,keepaspectratio,clip]{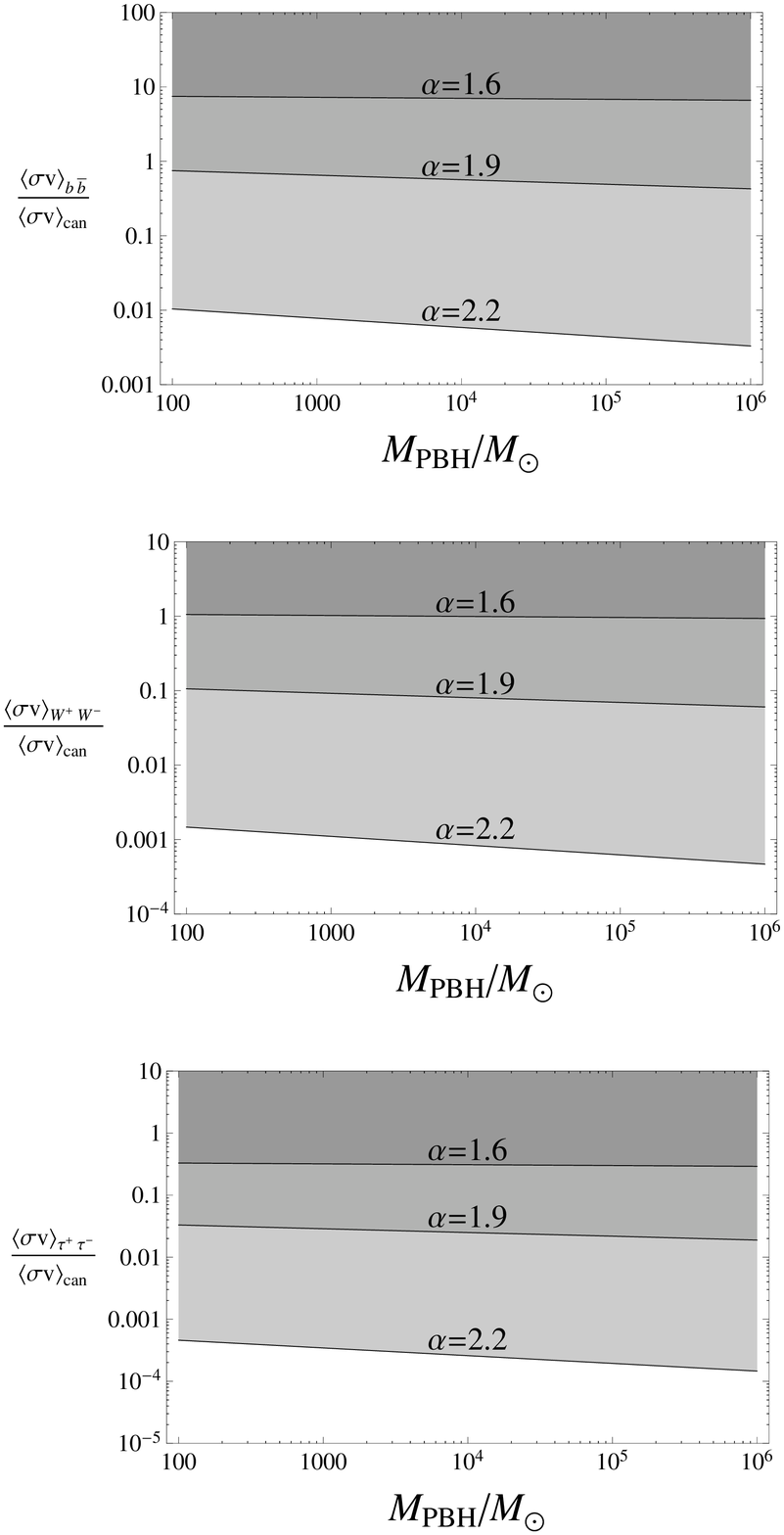}
\end{center}
\caption{
The constraint on the cross section of each mode as a function of $\MP$. 
For each value of $\alpha$, the shaded region is excluded. 
}
\label{nuconstmpbh}
\end{figure}
\begin{figure}[t]
\begin{center}
\includegraphics[width=11cm,keepaspectratio,clip]{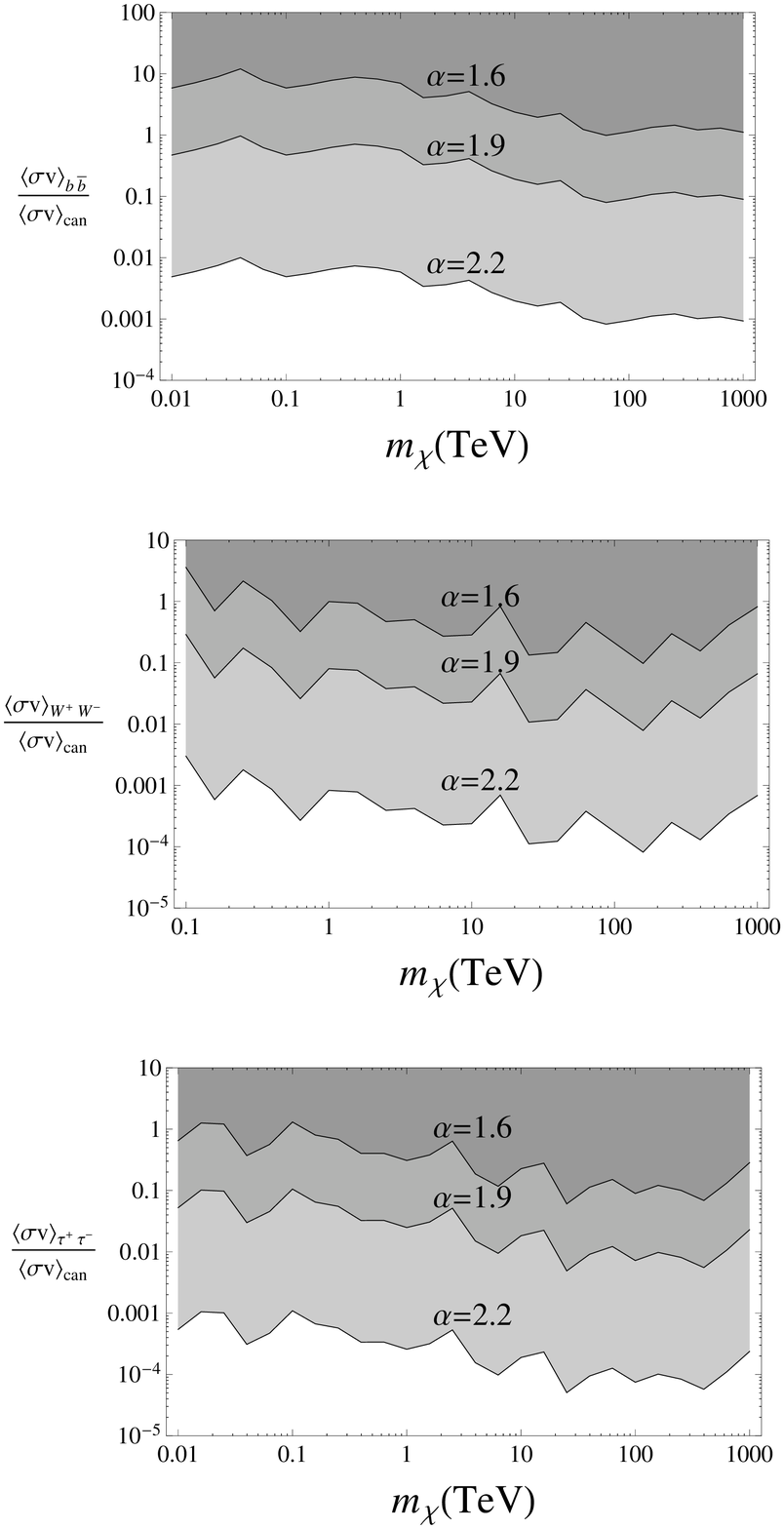}
\end{center}
\caption{
The dependence of the constraints of the cross sections on the dark matter mass.
}
\label{numdm}
\end{figure}

\section{Summary}
Super massive black holes (SMBHs) ($10^6 \sim 10^{9.5}~M_\odot$) are observed 
in many galaxies today as well as at high redshifts $z=6\sim 7$.
Furthermore, intermediate-mass black holes (IMBHs) ($10^2 \sim 10^6~M_\odot$) are also observed.
Since there is no established astrophysical explanation for their origin,
those massive black holes (MBHs) may be primordial black holes (PBHs). 

In this paper, we focused on the PBH scenario in which the MBHs (either SMBHs
or IMBHs, or both) are the remnants of the direct gravitational collapse of 
large density perturbation that occurred when the Universe is still dominated by radiation.
One of the attractive points of the PBH scenario is that we only need to assume large
amplitude of the seed density perturbation on scales corresponding to 
the Jeans length for the PBH formation.
Accordingly, several inflationary models have been contrived to realize large
density perturbation while satisfying the COBE normalization on cosmological scales.
This paper aims at discussing the possibility of excluding the PBH scenario
as the explanation of the observed MBHs.

In this paper, we have first considered the constraints on PBHs obtained from 
the CMB distortion that the seed density perturbation causes.
The PBH formation requires large amplitude of the density perturbation
and considerable amount of the CMB distortion is induced by the dissipation
of such density perturbation.
For the primordial power spectrum having a sharp peak at given scale, 
we found that PBHs in the mass range $M_{\rm PBH} \gtrsim 6\times 10^4~M_\odot$ 
are excluded as the seeds of the SMBHs.

Since PBHs lighter than $6 \times 10^4~M_\odot$ cannot be excluded from the
non-observation of the CMB distortion,
we have then proposed a new method which can potentially exclude smaller PBHs.
Because of fairly large density perturbation required for creating PBHs,
UCMHs are copiously produced as well (many more than the PBHs).
If DM is WIMP \cite{Jungman:1995df} having sizable interactions with SM particles, we expect 
cosmic rays such as $\gamma -$ rays and neutrinos out of UCMHs 
as a result of annihilations of DM particles.
If future terrestrial experiments identify DM particles as
WIMPs and measure the interaction strength between them and SM particles,
a situation can happen where the measured interaction strength is inconsistent with
non-observation of cosmic rays produced by DM annihilation inside the UCMHs,
the copious production of which is an inevitable outcome in the PBH scenario.
If such a situation actually arises, then our proposal disfavors the PBH
scenario and astrophysical explanations of the origin of the MBHs are strongly supported.
As far as we know, this is the second proposal that has the potential to falsify the
PBH scenario for a mass range $1~M_\odot \lesssim M_{\rm PBH} \lesssim 6 \times 10^4~M_\odot$, 
corresponding to the comoving wave number $3\times 10^{4}\mathrm{Mpc}^{-1}<k<7\times 10^6\mathrm{Mpc}^{-1}$. 
(The first proposal to test the PBH scenario is probably acoustic reheating\cite{Nakama:2014vla,Jeong:2014gna}.)

As typical annihilation modes, we considered three processes; $\chi \chi \to b{\bar b}$, 
$\chi \chi \to W^+W^-$ and $\chi \chi \to \tau^+\tau^-$, varying each annihilation cross section, 
and calculated the expected $\gamma -$ray and neutrino fluxes from the UCMHs assuming the PBH scenario.
By comparing them with the cosmic ray observations,
we found that the upper bounds on the cross sections for all the cases are 
several orders of magnitude smaller than the canonical value 
$\langle \sigma v \rangle_{\rm{can}} =3\times 10^{-26} {\rm cm}^3/{\rm s}$.
Since the WIMP in many extended SM models has annihilation cross sections to the SM particles
comparable to the canonical value \cite{Bertone:2004pz},
our bounds give severe restriction on the compatibility between the particle
physics models and the PBH scenario.
Our analysis clearly demonstrates that consideration of the cosmic rays out of the UCMHs 
could potentially falsify the PBH scenario.
If WIMPs having the standard interaction strength turn out to be DM in the future,
then our proposal will play an important role in elucidating one of the scenarios for the origin of the MBHs.

\section*{Acknowledgments}
We thank Takashi Hosokawa, Ryo Saito, Jun'ichi Yokoyama and Shuichiro Yokoyama for useful comments.
This work was supported
in part by the Grant-in-Aid for the Ministry of
Education, Culture, Sports, Science, and Technology,
Government of Japan, 
Nos. 21111006, 22244030,
23540327, 26105520 (K.K.), 
No.~25103505 (T.S.), 
No.~25.8199 (T.N.), 
and by the Center for the
Promotion of Integrated Science (CPIS) of Sokendai
(1HB5804100) (K.K.).
\bibliographystyle{unsrt}
\bibliography{ucmhref}

\begin{thebibliography}{10}

\bibitem{Kormendy:1995er}
John Kormendy and Douglas Richstone.
\newblock {Inward bound: The Search for supermassive black holes in galactic
  nuclei}.
\newblock {\em Ann.Rev.Astron.Astrophys.}, 33:581, 1995.

\bibitem{Fan:2003wd}
Xiaohui Fan et~al.
\newblock {A Survey of z $>$ 5.7 quasars in the Sloan Digital Sky Survey. 2.
  Discovery of three additional quasars at z $>$ 6}.
\newblock {\em Astron.J.}, 125:1649, 2003.

\bibitem{Mortlock:2011va}
Daniel~J. Mortlock, Stephen~J. Warren, Bram~P. Venemans, Mitesh Patel, Paul~C.
  Hewett, et~al.
\newblock {A luminous quasar at a redshift of z = 7.085}.
\newblock {\em Nature}, 474:616, 2011.

\bibitem{Maccarone:2007dd}
Thomas~J. Maccarone, Arunav Kundu, Stephen~E. Zepf, and Katherine~L. Rhode.
\newblock {A black hole in a globular cluster}.
\newblock {\em Nature}, 445:183--185, 2007.

\bibitem{Farrell:2010bf}
Sean Farrell, Natalie Webb, Didier Barret, Olivier Godet, and Joana Rodrigues.
\newblock {An Intermediate-mass Black Hole of Over 500 Solar Masses in the
  Galaxy ESO 243-49}.
\newblock {\em Nature}, 460:73--75, 2009.

\bibitem{Yokoyama:1995ex}
Junichi Yokoyama.
\newblock {Formation of MACHO primordial black holes in inflationary
  cosmology}.
\newblock {\em Astron.Astrophys.}, 318:673, 1997.

\bibitem{Duechting:2004dk}
Norbert Duechting.
\newblock {Supermassive black holes from primordial black hole seeds}.
\newblock {\em Phys.Rev.}, D70:064015, 2004.

\bibitem{Kawasaki:2012kn}
Masahiro Kawasaki, Alexander Kusenko, and Tsutomu~T. Yanagida.
\newblock {Primordial seeds of supermassive black holes}.
\newblock {\em Phys.Lett.}, B711:1--5, 2012.

\bibitem{Kawasaki:2012wr}
Masahiro Kawasaki, Naoya Kitajima, and Tsutomu~T. Yanagida.
\newblock {Primordial black hole formation from an axion-like curvaton model}.
\newblock {\em Phys.Rev.}, D87:063519, 2013.

\bibitem{Kohri:2012yw}
Kazunori Kohri, Chia-Min Lin, and Tomohiro Matsuda.
\newblock {Primordial black holes from the inflating curvaton}.
\newblock {\em Phys.Rev.}, D87(10):103527, 2013.

\bibitem{Hawking:1971ei}
Stephen Hawking.
\newblock {Gravitationally collapsed objects of very low mass}.
\newblock {\em Mon.Not.Roy.Astron.Soc.}, 152:75, 1971.

\bibitem{Carr:1974nx}
Bernard~J. Carr and S.W. Hawking.
\newblock {Black holes in the early Universe}.
\newblock {\em Mon.Not.Roy.Astron.Soc.}, 168:399--415, 1974.

\bibitem{Carr:1975qj}
Bernard~J. Carr.
\newblock {The Primordial black hole mass spectrum}.
\newblock {\em Astrophys.J.}, 201:1--19, 1975.

\bibitem{Carr:2009jm}
B.J. Carr, Kazunori Kohri, Yuuiti Sendouda, and Jun'ichi Yokoyama.
\newblock {New cosmological constraints on primordial black holes}.
\newblock {\em Phys.Rev.}, D81:104019, 2010.

\bibitem{Kohri:2007qn}
Kazunori Kohri, David~H. Lyth, and Alessandro Melchiorri.
\newblock {Black hole formation and slow-roll inflation}.
\newblock {\em JCAP}, 0804:038, 2008.

\bibitem{Alabidi:2012ex}
Laila Alabidi, Kazunori Kohri, Misao Sasaki, and Yuuiti Sendouda.
\newblock {Observable Spectra of Induced Gravitational Waves from Inflation}.
\newblock {\em JCAP}, 1209:017, 2012.

\bibitem{Carr:1993aq}
Bernard~J. Carr and James~E. Lidsey.
\newblock {Primordial black holes and generalized constraints on chaotic
  inflation}.
\newblock {\em Phys.Rev.}, D48:543--553, 1993.

\bibitem{Carr:1994ar}
Bernard~J. Carr, J.H. Gilbert, and James~E. Lidsey.
\newblock {Black hole relics and inflation: Limits on blue perturbation
  spectra}.
\newblock {\em Phys.Rev.}, D50:4853--4867, 1994.

\bibitem{Fixsen:1996nj}
D.J. Fixsen, E.S. Cheng, J.M. Gales, John~C. Mather, R.A. Shafer, et~al.
\newblock {The Cosmic Microwave Background spectrum from the full COBE FIRAS
  data set}.
\newblock {\em Astrophys.J.}, 473:576, 1996.

\bibitem{Chluba:2012we}
Jens Chluba, Adrienne~L. Erickcek, and Ido Ben-Dayan.
\newblock {Probing the inflaton: Small-scale power spectrum constraints from
  measurements of the CMB energy spectrum}.
\newblock {\em Astrophys.J.}, 758:76, 2012.

\bibitem{Bean:2002kx}
Rachel Bean and Joao Magueijo.
\newblock {Could supermassive black holes be quintessential primordial black
  holes?}
\newblock {\em Phys.Rev.}, D66:063505, 2002.

\bibitem{Ricotti:2009bs}
M.~Ricotti and A.~Gould.
\newblock {A New Probe of Dark Matter and High-Energy Universe Using
  Microlensing}.
\newblock {\em Astrophys.J.}, 707:979--987, 2009.

\bibitem{Scott:2009tu}
Pat Scott and Sofia Sivertsson.
\newblock {Gamma-Rays from Ultracompact Primordial Dark Matter Minihalos}.
\newblock {\em Phys.Rev.Lett.}, 103:211301, 2009.

\bibitem{Lacki:2010zf}
Brian~C. Lacki and John~F. Beacom.
\newblock {Primordial Black Holes as Dark Matter: Almost All or Almost
  Nothing}.
\newblock {\em Astrophys.J.}, 720:L67--L71, 2010.

\bibitem{Berezinsky:2010kq}
V.~Berezinsky, V.~Dokuchaev, Yu. Eroshenko, M.~Kachelriess, and M.~Aa. Solberg.
\newblock {Superdense cosmological dark matter clumps}.
\newblock {\em Phys.Rev.}, D81:103529, 2010.

\bibitem{Josan:2010vn}
Amandeep~S. Josan and Anne~M. Green.
\newblock {Gamma-rays from ultracompact minihalos: potential constraints on the
  primordial curvature perturbation}.
\newblock {\em Phys.Rev.}, D82:083527, 2010.

\bibitem{Yang:2011jb}
Yupeng Yang, Xiaoyuan Huang, Xuelei Chen, and Hongshi Zong.
\newblock {New Constraints on Primordial Minihalo Abundance Using Cosmic
  Microwave Background Observations}.
\newblock {\em Phys.Rev.}, D84:043506, 2011.

\bibitem{Bringmann:2011ut}
Torsten Bringmann, Pat Scott, and Yashar Akrami.
\newblock {Improved constraints on the primordial power spectrum at small
  scales from ultracompact minihalos}.
\newblock {\em Phys.Rev.}, D85:125027, 2012.

\bibitem{Yang:2011eg}
Yu-Peng Yang, Lei Feng, Xiao-Yuan Huang, Xuelei Chen, Tan Lu, et~al.
\newblock {Constraints on ultracompact minihalos from extragalactic
  $\gamma$-ray background}.
\newblock {\em JCAP}, 1112:020, 2011.

\bibitem{Li:2012qha}
Fangda Li, Adrienne~L. Erickcek, and Nicholas~M. Law.
\newblock {A new probe of the small-scale primordial power spectrum:
  astrometric microlensing by ultracompact minihalos}.
\newblock {\em Phys.Rev.}, D86:043519, 2012.

\bibitem{Yang:2012qi}
Yupeng Yang, Guilin Yang, Xiaoyuan Huang, Xuelei Chen, Tan Lu, et~al.
\newblock {The contribution of ultracompact dark matter minihalos to the
  isotropic radio background}.
\newblock {\em Phys.Rev.}, D87:083519, 2012.

\bibitem{Yang:2013dsa}
Yupeng Yang, Guilin Yang, and Hongshi Zong.
\newblock {Neutrino signals from ultracompact minihalos and constraints on the
  primordial curvature perturbation}.
\newblock {\em Phys. Rev.}, D87:103525, 2013.

\bibitem{Sjostrand:2006za}
Torbjorn Sjostrand, Stephen Mrenna, and Peter~Z. Skands.
\newblock {PYTHIA 6.4 Physics and Manual}.
\newblock {\em JHEP}, 0605:026, 2006.

\bibitem{Josan:2009qn}
Amandeep~S. Josan, Anne~M. Green, and Karim~A. Malik.
\newblock {Generalised constraints on the curvature perturbation from
  primordial black holes}.
\newblock {\em Phys.Rev.}, D79:103520, 2009.

\bibitem{Kelly:2011ab}
Brandon~C. Kelly and Andrea Merloni.
\newblock {Mass Functions of Supermassive Black Holes Across Cosmic Time}.
\newblock {\em Adv.Astron.}, 2012:970858, 2012.

\bibitem{Kim:1996hr}
Hee~Il Kim and Chul~H. Lee.
\newblock {Constraints on the spectral index from primordial black holes}.
\newblock {\em Phys.Rev.}, D54:6001--6007, 1996.

\bibitem{Shibata:1999zs}
Masaru Shibata and Misao Sasaki.
\newblock {Black hole formation in the Friedmann universe: Formulation and
  computation in numerical relativity}.
\newblock {\em Phys.Rev.}, D60:084002, 1999.

\bibitem{Polnarev:2006aa}
Alexander~G. Polnarev and Ilia Musco.
\newblock {Curvature profiles as initial conditions for primordial black hole
  formation}.
\newblock {\em Class.Quant.Grav.}, 24:1405--1432, 2007.

\bibitem{Hidalgo:2008mv}
J.C. Hidalgo and A.G. Polnarev.
\newblock {Probability of primordial black hole formation and its dependence on
  the radial profile of initial configurations}.
\newblock {\em Phys.Rev.}, D79:044006, 2009.

\bibitem{1475-7516-2014-01-037}
Tomohiro Nakama, Tomohiro Harada, A.G. Polnarev, and Jun'ichi Yokoyama.
\newblock Identifying the most crucial parameters of the initial curvature
  profile for primordial black hole formation.
\newblock {\em JCAP}, 2014(01):037, 2014.

\bibitem{Green:2005fa}
Anne~M. Green, Stefan Hofmann, and Dominik~J. Schwarz.
\newblock {The First wimpy halos}.
\newblock {\em JCAP}, 0508:003, 2005.

\bibitem{Berezinsky:2007qu}
Veniamin Berezinsky, Vyacheslav Dokuchaev, and Yury Eroshenko.
\newblock {Remnants of dark matter clumps}.
\newblock {\em Phys.Rev.}, D77:083519, 2008.

\bibitem{Fillmore:1984wk}
J.A. Fillmore and P.~Goldreich.
\newblock {Self-similiar gravitational collapse in an expanding universe}.
\newblock {\em Astrophys.J.}, 281:1--8, 1984.

\bibitem{Bertschinger:1985pd}
E.~Bertschinger.
\newblock {Self - similar secondary infall and accretion in an Einstein-de
  Sitter universe}.
\newblock {\em Astrophys.J.Suppl.}, 58:39, 1985.

\bibitem{Ricotti:2007au}
Massimo Ricotti, Jeremiah~P. Ostriker, and Katherine~J. Mack.
\newblock {Effect of Primordial Black Holes on the Cosmic Microwave Background
  and Cosmological Parameter Estimates}.
\newblock {\em Astrophys.J.}, 680:829--845, 2008.

\bibitem{Battaglia:2005rj}
Giuseppina Battaglia, Amina Helmi, Heather Morrison, Paul Harding, Edward~W.
  Olszewski, et~al.
\newblock {The Radial velocity dispersion profile of the Galactic Halo:
  Constraining the density profile of the dark halo of the Milky Way}.
\newblock {\em Mon.Not.Roy.Astron.Soc.}, 364:433--442, 2005.

\bibitem{Abdo:2010nz}
A.A. Abdo et~al.
\newblock {The Spectrum of the Isotropic Diffuse Gamma-Ray Emission Derived
  From First-Year Fermi Large Area Telescope Data}.
\newblock {\em Phys.Rev.Lett.}, 104:101101, 2010.

\bibitem{Lee:2012pz}
Fei-Fan Lee, Guey-Lin Lin, and Yue-Lin~Sming Tsai.
\newblock {Sensitivities of IceCube DeepCore Detector to Signatures of Low-Mass
  Dark Matter in the Galactic Halo}.
\newblock {\em Phys.Rev.}, D87:025003, 2013.

\bibitem{Jungman:1995df}
Gerard Jungman, Marc Kamionkowski, and Kim Griest.
\newblock {Supersymmetric dark matter}.
\newblock {\em Phys.Rept.}, 267:195--373, 1996.

\bibitem{Nakama:2014vla}
Tomohiro Nakama, Teruaki Suyama, and Jun'ichi Yokoyama.
\newblock {Reheating the Universe Once More: The Dissipation of Acoustic Waves
  as a Novel Probe of Primordial Inhomogeneities on Even Smaller Scales}.
\newblock {\em Phys.Rev.Lett.}, 113:061302, 2014.

\bibitem{Jeong:2014gna}
Donghui Jeong, Josef Pradler, Jens Chluba, and Marc Kamionkowski.
\newblock {Silk damping at a redshift of a billion: a new limit on small-scale
  adiabatic perturbations}.
\newblock {\em Phys.Rev.Lett.}, 113:061301, 2014.

\bibitem{Bertone:2004pz}
Gianfranco Bertone, Dan Hooper, and Joseph Silk.
\newblock {Particle dark matter: Evidence, candidates and constraints}.
\newblock {\em Phys.Rept.}, 405:279--390, 2005.

\end{thebibliography}
\end{document}